\documentclass{article}
\usepackage{amsmath}
\usepackage{cite}
\usepackage{tikz}
\usepackage{tikz,tkz-tab}
\usepackage{amsfonts}%
\usepackage{amssymb}%
\usepackage{hyperref}
\usepackage{mathtools}
\usepackage{color}
\usepackage{setspace}
\usepackage{empheq}
\usepackage{bbm, dsfont}
\usepackage{dsfont}
\usepackage{mathtools}
\usepackage{geometry}
\usepackage{enumitem} 
\usepackage[bottom]{footmisc}
\usetikzlibrary{arrows}
\usepackage{lscape}
\usepackage{tcolorbox}
\usepackage{caption}
\usepackage{subfig}
\usepackage{hyperref}

\newtheorem{definition}{Definition}

\usetikzlibrary{positioning}
\tikzset{main node/.style={circle,draw,minimum size=0.5cm,inner sep=0pt},
            }

\newtheorem{example}{Example}

\newtheorem{proposition}{Proposition}

\begin{document}

\title{Targeting in Social Networks with Anonymized Information}

\author{Francis Bloch\footnote{Paris School of Economics, Universit\'{e} Paris1 Panth\'{e}on-Sorbonne, 48 Boulevard Jourdan, 75014 Paris, France. e-mail:  francis.bloch@univ-paris1.fr}  \and Shaden Shabayek\footnote{Paris School of Economics, Universit\'{e} Paris1 Panth\'{e}on-Sorbonne, 48 Boulevard Jourdan, 75014 Paris, France.e-mail:shaden.shabayek@psemail.eu} }
\maketitle

\begin{abstract} This paper studies whether a planner who only has information about the network topology can discriminate among agents according to their network position. The planner proposes a simple menu of contracts, one for each location, in order to maximize total welfare, and agents choose among the menu. This mechanism is immune to deviations by single agents, and to deviations by groups of agents of sizes $2$, $3$ and $4$ if side-payments are ruled out. However, if compensations are allowed, groups of agents may have an incentive to jointly deviate from the optimal contract in order to exploit other agents. We identify network topologies for which the optimal contract is group incentive compatible with transfers: undirected networks and regular oriented trees, and network topologies for which the planner must assign uniform quantities: single root and nested neighborhoods directed networks.
 \end{abstract}
 
 \bigskip

\noindent {\sc Keywords}: social networks, targeting, privacy protection, mechanism design.

\medskip

\noindent {\sc JEL Classification numbers}: D85, D82, D42,L14.

\onehalfspace

\newpage
\section{Introduction}

For privacy and security reasons, individual data are often anonymized before they can be shared or sold to third parties. As is known from the statistical and computer science literature, privacy-preserving data disclosure can be achieved in multiple ways, from a complete elimination of any identification of the individual (anonymization) to algorithmic and statistical methods to avoid identification (differential privacy). The recent advances in privacy protection, together with the fast development of data sharing and data disclosure raises the following question: How valuable are anonymized data (rather than full individual data) for a third party whose objective is not aligned with the welfare of individuals? When will the third party be able to achieve the same value with anonymized data and with the full, individual data?

We answer this question in the specific context of {\em social network data}. In that context,  anonymized data describe the {\em network topology}, the architecture of the network without specifying the identity of agents at different nodes.  For example, anonymized social network data can be geo-data describing a network of roads and houses with no specification of the inhabitants, or an organizational chart for a company or a criminal organization with no identification of the individuals, or  a snapshot of a fraction of a large digital social network (Twitter, LinkedIn or Facebook) with lacunary evidence on the identity of the nodes. 

We suppose that the social network describes local positive externalities in the choices of agents, and consider a third party whose objective is to maximize total surplus. The third party could be a monopolist, selecting discriminatory prices to exploit local consumption externalities (by lowering the price to stimulate consumption of individuals generating important externalities), or a firm who wishes to maximize output, when the social networks describes synergies in efforts by pairs of workers. In the presence of local network externalities, the objective of the planner and of individuals are not aligned: the planner internalizes positive externalities that an agent produces on other agents, raising consumption (in the case of the monopolist) or effort (in the case of the firm) beyond the optimum of individuals.

When full individual data are available, the planner can target agents and offer them the exact contract which maximizes her objective. But when the planner only knows the network topology, she is only able to offer a contract for each {\em location} in the network, with no certainty that agents will indeed choose the contract corresponding to that location. This is the incentive problem we study in this paper, where we analyze for which network architectures this incentive constraint is binding or not. 

We first observe, unsurprisingly, that correlation in the locations of agents allows the planner to achieve his first-best outcome when agents can only make unilateral deviations. Indeed, because the set of locations in the network is fixed and finite, if the planner proposes a menu of contracts, one for each location, and all other agents pick the contract corresponding to their location, the deviating agent must be selecting the same contract as another agent. The planner can use this evidence to ascertain that one of the agents has lied, and punish all agents (for example by offering a level lower than any agent's optimum) as in Cr\'{e}mer and McLean   \cite{cremer1988full}, thereby ensuring that all agents have an incentive to choose the contract corresponding to their true location.\footnote{As usual, this reasoning only shows that truth-telling is one equilibrium of the revelation game played by the agents, but other equilibria exist as well -- in fact, in this model, {\em any} choice of contracts by the agents is an equilibrium of the game.}

We thus turn our attention to joint deviations, where subsets of agents decide to ``lie'' about their locations, and choose contracts which do not correspond to their true position in the network. Given that the objective of the planner is to maximize social surplus, why do groups of agents have an incentive to lie about their locations and exchange their contracts? Why would an agent who already receives a consumption or effort level above his optimal choice accept to exchange his contract with another agent with a higher level? The answer to these questions stems from the fact that, by exchanging their contracts, agents are able to increase the externalities they receive from other agents, while reducing the externalities they create for agents outside the deviating coalition. For example, consider a directed network architecture where the hub of the star (the ``influencer") produces positive consumption externalities on all other agents but receives no externalities in return. The planner, internalizing this externality, assigns a very high consumption to the hub, and a lower consumption to a peripheral agents. But the hub may have an incentive to exchange his contract with a peripheral agent, increasing his utility by reducing his consumption level, at the expense of all other peripheral agents, who now receive a much lower level of externalities.

The first main Proposition of the paper asserts that, when the size of the deviating group is two, three or four, the optimal contract is in fact immune to deviations by groups of agents. The proof of the Proposition in the case of pairwise deviations is clear. For an agent to accept to exchange his contract for a higher one  (e.g. for a peripheral agent to accept the high level of the hub), whereas his current contract is already too high given the contracts of other agents, he must receive higher externalities in return from the members of the deviating coalition. But when a pair of agents deviates, the externalities inside the deviating coalition do not change, as agents only exchange their contracts and externalities are suppose to be symmetric. Hence, in any pairwise deviation, absent any compensation, the agent with the lower contract must be unwilling to exchange his contract with the other agent.\footnote{Unfortunately, this intuition does not easily carry over to deviations by larger groups, where agents can benefit from a change in the level of externalities among agents in the deviating coalition. For groups of three and four agents, we are able to prove that one of the lower contract agents must be unwilling to exchange, but so far have been unable to generalize the result beyond groups of four agents.}

Based on this observation, we consider a more restrictive group incentive compatibility condition, by allowing agents in the deviating coalition to make side-payments. Under this condition, a contract is group incentive compatible if no coalition, by choosing a permutation of the locations, can increase the sum of utilities of its members. In the particular context of a linear-quadratic game, we show that this condition is equivalent to assuming that, whenever an agent $i$ has a higher contract than an agent $j$, the sum of externalities received by agent $i$ must be at least as large as the sum of externalities received by agent $j$.

This characterization drives the second main result of the paper. When externalities are reciprocal (the graph representing the social network is undirected), then the optimal contract of the planner is immune to group deviations with transfers. The intuition underlying this result is that an agent will receive a higher contract  in the first-best if and only if he is more central (in the sense of Katz-Bonacich) than the other agent. But this centrality measure is defined recursively: an agent has a higher Katz-Bonacich centrality if and only if the sum of the Katz-Bonacich centralities of his neighbors is higher. In other words, when influence is reciprocal, an agent with a higher contract in the first-best is also an agent who receives (and produces) higher externalities, and the sum of utilities of the agents cannot increase through an exchange of contracts.

The situation changes dramatically when externalities are asymmetric. We next focus on directed networks with a hierarchical structure: the set of agents can be partitioned into an ordered collection of subsets (or tiers of the hierarchy) such that agents at higher tiers influence agents at lower tiers, but agents at lower tiers never influence agents at higher tiers. In this family of hierarchical social networks, we first show that the optimal contract is group incentive compatible when side-payments are ruled out. But when agents in a deviating coalition can make transfers, we uncover two situations where the optimal contract of the planner cannot be sustained. 

First, when there exists one agent who influences all other agents {\em the single root} and the number of tiers is larger than two, then in an optimal contract which is immune to group deviations with transfers, {\em all agents in tiers lower than two must receive the same contract as the root}. The intuition is as follows: if the root receives a higher contract than agents at lower tiers, there is an incentive to exchange the higher quantity of the root with the lower quantity of the lower tier agent, as the latter receives externalities but not the former. Hence, any contract which is immune to deviations by groups with transfers must assign a lower contract to the root than to agents at tiers lower than two.\footnote{For agents at tier two, the only externalities they receive come from the root, and hence there is no restriction on the relative ranking of their contract with the contract of the root.} However, as the root produces externalities on all other agents, whereas other agents may not influence or be influenced by all other agents, the planner also has an incentive to increase the contract of the root with respect to the contract of any other agent. Hence, at the optimum, the contract of the root must be exactly equal to the contract offered to all agents at tiers lower than two. Notice that this implies that targeting is impossible, and may result in large efficiency losses with respect to the first-best.

Second, suppose that {\em neighborhoods are nested}, so that an agent at a lower tier is influenced by all agents who also influence agents who influence him. Multiple roots can exist, and agents can be connected to multiple roots, but the structure of nested neighborhoods implies that the set of influencers of an agent is included in the set of influencers of the agent he influences. Assume furthermore that the size of the tiers is increasing: there are strictly more agents at lower tiers. In this hierarchical structure, as in the case of single roots, we show that the optimal contract is not group incentive compatible with transfers, and that any root has an incentive to exchange his contract with an agent at lower tier. As in the case of a single root, the planner would like to increase the quantity of the root, so that {\em the optimal contract which is group incentive compatible with transfers results in a uniform contract for all agents.}\footnote{This time, under the assumption that agents at tier two are influenced by more than one root, they receive more externalities than a root, and hence the reasoning which implies that an agent must receive a higher contract than the root in a group incentive compatible contract with transfers also holds for tier two agents.} Again, the group incentive compatibility constraint is binding, and induces a total inability to discriminate among agents.

Finally, we consider a hierarchical structure where the optimal contract is immune to deviations by coalitions making side payments: {\em regular oriented tree}. In these structures, every agent is influenced by a single other agent (the social network is an oriented tree). Furthermore, the number of agents influenced by an agent at any given tier is identical across agents of the same tier, and strictly decreasing with the tier (this is what we call a ``regular" oriented tree). In that specific hierarchy, agents at higher tier must have a higher contract than the agents they influence, because every agent they influence receives his externalities from his immediate predecessor, but the predecessor receives externalities from other agents). When the structure is regular, the planner will also want agents at higher levels of the hierarchy to receive higher contracts, so that the first-best contract is indeed group incentive compatible with transfers.

Our results thus show that the optimal contract of the planner is immune to deviations by groups whenever influence is reciprocal, or agents at higher levels of the hierarchy receive more influence than agents at lower levels of the hierarchy. When on the other hand agents at higher levels of the hierarchy produce more influence but receive less influence than agents at lower tiers, the first-best contract is susceptible to group deviations, and the planner may in extreme cases be forced to assign a uniform contract to all agents. These results thus show how the architecture of the network matters to determine whether  anonymized data are sufficient or not for the planner to implement the first-best.

We also explore the robustness of our results with respect to the assumptions made in the model. First, we note that we put a very strong restriction on the structure of deviating coalitions by assuming that they are adjacent (i.e. any pair of agents in the deviating coalition must be connected by a link). Absent this restriction, more coalitions can deviate and in fact, even in the case of undirected networks and regular oriented trees, the first-best contract may become impossible to implement. 

Second, we analyze the effect of increasing the complexity of the contract, by allowing the planner to extract information about agents' local neighborhood. We show that if the planner can collect information about an agent's neighbors' identity, the first-best contract becomes immune to group deviations with transfers. Finally, we explore what happens when the planner knows the identity of some agents in the network. For example, suppose that the network knows the identity of the main influencers (the roots of the directed social network). We show through an example that this partial information may greatly help the planner and allow him to discriminate much more effectively among the agents.

\vskip 0.2cm
\noindent \textbf{Related literature.}  Our paper is related to the study of targeting in social networks, as surveyed in  Bloch  \cite{bloch2016targeting}. The linear-quadratic model of interaction in networks we consider was introduced in Ballester, Calvo-Armengol and Zenou  \cite{ballester2006s}. General problems of targeting where the planner seeks to maximize social welfare in the presence of  complementarities have been recently studied by Demange  \cite{demange2017optimal} and Galeotti, Golub and Goyal\cite{galeotti2017targeting}.

The problem of a monopolist pricing in the presence of local consumption externalities has first been studied by  Candogan, Bimpikis and Ozdaglar \cite{candogan2012optimal} and Bloch and Qu\'{e}rou  \cite{bloch2013pricing}. Candogan, Bimpikis and Ozdaglar  \cite{candogan2012optimal} and Bloch and Qu\'{e}rou  \cite{bloch2013pricing} both look at the relation between the centrality of consumers in a given network and the prices and quantities they are offered by a monopolist; in the context of perfect knowledge of the network structure. They both show that when the network is undirected - that is two directly connected consumers influence each other equally -  consumers are offered quantities proportional to their Bonacich centrality for the same price. Fainmesser and Galeotti \cite{fainmesser2015pricing} consider a monopolist which can price discriminate based on her knowledge of either the in-degrees of consumers (their influence), the out-degree of consumers (their susceptibility to influence) or both. They show that the knowledge of in-degree (respectively out-degree) is more valuable when the dispersion in in-degrees (respectively out-degree) in the network is higher.

A handful of recent papers extend the analysis of monopoly pricing  and targeting to situations of incomplete information. Jadbabaie and Kakhbod  \cite{jadbabaie2019optimal} consider the pricing problem when the strength of influence among agents is private information. Ata, Belloni and Candogan  \cite{ata2018latent}, take a different approach and assume that there is a  seller who faces two groups of agents: observable and latent. There are local consumption externalities among members of the same group and across groups. The seller can observe past purchasing decisions only of the group of $observable$ agents and can try to deduce from these observations information on influence patterns. Finally, in a paper which shares our motivation about the planner's incentives to elicit information about the network, Shi and Xing \cite{shi2018screening} consider a model where buyers draw their in and out-degrees independently from a common distribution. They characterize the optimal contracts in the framework of random graphs. The novelty in our model is to introduce incomplete information with regard to the network itself by assuming that the structure is observable but the locations or identities of agents is private information.

\section{The model} \label{themodel}

\subsection{Agents, utilities and network effects}

We consider a set $N$ of agents (labeled ``he"), indexed by $i=1,2,..,n$. Agents are connected in a social network, which we represent by a graph $g$, with nodes $i=1,2,..n$ and edges $g_{ij} \in \{0,1\}$. We denote by ${\bf G}$ the adjacency matrix of the graph $g$. 

Each agent $i$ chooses an action $x_i \in \Re_+$, which will be interpreted as his consumption or effort. The utility of agent $i$ depends on his action and the actions of his neighbors in the social network. Following a well-established model, initiated by Ballester, Calvo-Armengol and Zenou \cite{ballester2006s}, we assume that utilities are quadratic, so that 

\[ U_i = a_i  x_i - \frac{b_i}{2} x_i^2 + \alpha \sum_{j \in N} g_{ij} x_i x_j.\]

For the main part of the analysis, we assume that all agents are identical, except for their location in the social network. For homogeneous agents, $a_i=a$ and we normalize  $b_i=b=1$ so that

\[ U_i = a x_i - \frac{1}{2} x_i^2 + \alpha \sum_{j \in N} g_{ij} x_i x_j.\]


\subsection{The planner's first-best} 

We consider a planner (labeled ``she") with perfect knowledge of the network $g$ who chooses the vector of actions ${\bf x} = (x_1,..x_n)$ in order to maximize the sum of utilities of the agents,

\begin{equation}
\label{objective}
V = \sum_i U_i = \sum_i a x_i - \frac{1}{2} x_i^2 + \alpha \sum_j g_{ij} x_i x_j. 
\end{equation}

The first-order conditions result in a system of linear equations:

\[ a-  x_i + \alpha \sum_{j \in N} (g_{ij} + g_{ji}) x_j = 0, \]

\noindent or in matrix terms, letting ${\bf 1 }$ denote the vector $(1,...,n)$, ${\bf 0}$ the vector $(0,..,0)$ and ${\bf I}$ the identity matrix

\[ a{\bf 1 } - {\bf I} {\bf x} + \alpha ({\bf G} +{ \bf G}^T) {\bf x} = {\bf 0}.\]

Let $\lambda$ denote the largest eigenvalue of the matrix $({\bf G} +{ \bf G}^T)$.\footnote{Because the matrix $({\bf G} +{ \bf G}^T)$ is symmetric, all its eigenvalues are real and hence the largest eigenvalue is well-defined.}  Following Ballester, Calvo-Armengol and Zenou \cite{ballester2006s} we compute the optimal solution of the planner's problem as the solution to the system of linear equations when external effects are not too large,

\begin{proposition} \label{propzero} If $1 > \alpha \lambda$ the optimal solution of the planner 's problem is given by
\[ {\bf x}^* = ( {\bf I} - \alpha ({\bf G} +{ \bf G}^T))^{-1} a {\bf 1}. \]
\end{proposition}

Next  recall that the {\em Katz-Bonacich} centrality measure (Katz \cite{katz1953new} and Bonacich \cite{bonacich1972factoring}) of an agent in the network with $g$ with  discount factor $\delta$, $\beta({\bf G}, \delta) $ is the discounted sum of walks originating from that agent,

\begin{eqnarray*}
 \beta({\bf G}, \delta) & = &  \sum_{k=0}^{\infty}  \delta^k {\bf G}^k {\bf 1}, \\
 & = & ({\bf I} - \delta {\bf G})^{-1} {\bf 1}.
 \end{eqnarray*}
 
 It is easy to check that , the planner's solution is to assign to each agent $i$ an action which is proportional to its Katz-Bonacich centrality measure in the network $({\bf G} + {\bf G})^T$ under discount factor $\alpha$,
 
 \[ {\bf x}^* \propto \beta(({\bf G} + {\bf G}^T), \alpha). \]
 
Hence the optimal choice of the planner is to assign discriminatory actions to the agents, in proportion to their Bonacich centrality measure in the network $({\bf G} + {\bf G})^T$.\footnote{Strictly speaking, $({\bf G} + {\bf G})^T$ is a weighted network, with weights equal to $0, 1$ or $2$ on each link.} We now observe that the planner's general problem encompasses different situations of pricing and targeting in networks.

\subsubsection{Pricing with network externalities} Consider, as in Candogan, Bimpikis and Ozdaglar  \cite{candogan2012optimal} and Bloch and Qu\'{e}rou  \cite{bloch2013pricing} a monopolist setting prices on a market where consumers experience positive consumption externalities. For simplicity, assume that the monopolist faces a constant marginal unit cost $c$. Consumer $i$'s utility depends both on his own consumption and on the consumption of his neighbors in the social network. If the monopolist sets a unit price $p_i$ to consumer $i$, and offers a quantity $x_i$, the utility of the consumer is given by

\[ U_i = a x_i - \frac{1}{2} x_i^2 + \alpha \sum_{j \in N} g_{ij} x_i x_j - p_i x_i.\]

Assuming that the monopolist can fully discriminate among consumers, she will select a price $p_i$ to capture the entire surplus of the consumer and make a profit 

\[ \pi = \sum_{i \in N} (p_i -c)x_i  = \sum_{i \in N} \big( (a-c) x_i - \frac{1}{2} x_i^2 + \alpha \sum_{j \in N} g_{ij} x_i x_j \big) , \]

\noindent an objective function which is  equivalent to the objective function $V$ in equation (\ref{objective}) after a renormalization.

\subsubsection{Targeting and efforts} Next we consider as in Galeotti, Golub and Goyal   \cite{galeotti2017targeting} the decision $x_i$  chosen by an agent with utility function

\[ U_i = a  x_i - \frac{1}{2} x_i^2 + \alpha \sum_{j \in N} g_{ij} x_i x_j.\]

The planner intervenes by subsidizing or taxing the marginal stand-alone utility of the effort, resulting in $\hat{a}_i = a +t_i$, where $t_i$ is the tax or subsidy of agent $i$. Any intervention leading to a change from $a$ to $\hat{a}_i$ will result in a change in the effort level $x_i$.  Hence an intervention can be interpreted as a choice of decisions $x_i$ to maximize the sum of utilities of the agents

\[ V = \sum_{i \in N} \big( a x_i - \frac{1}{2} x_i^2 + \alpha \sum_{j \in N} g_{ij} x_i x_j \big).\footnote{Note that Galeotti, Golub and Goyal   \cite{galeotti2017targeting}  assume the planner faces a fixed budget and incurs a cost which is quadratic in the difference between the initial value $a$ and the target value $\hat{a_i}$. This results in an additional constraint in the planner's problem. When the budget is sufficiently large, this additional constraint is not binding and the problem is equivalent to the problem we consider here.}\]

\subsection{Network architectures and information structures}

Proposition \ref{propzero} characterizes the first-best solution when the planner has complete information on the social network. We now introduce the constraints faced by the planner when she only has access to {\em anonymized} information on the network architecture. 

Consider two networks $g$ and $g'$. These networks share the same {\em network architecture} if they are equivalent, up to a relabeling of the agents. Formally, two networks $g$ and $g'$ are equivalent (denoted $g E g'$)  if  there exists a permutation $\pi$ of the nodes such that $g'_{ij} = g_{\pi(i) \pi(j)}$ for all $i,j$. A {\em network architecture}  $\hat{g}$ is an equivalence class of the relation $E$ over the set of all graphs. The cardinality of the equivalence class $\hat{g}$ varies according to the graph $g$ and depends on the symmetry group of the graph $g$.  If the graph is symmetric (for example an empty graph or a complete graph), there is exactly one element in $\hat{g}$ and the information of the planner is complete. If on the other hand the symmetry group of the graph $g$ is empty and all nodes occupy different positions, there are exactly $n!$ elements in the  equivalence class $\hat{g}$. If there exists a subset of nodes $S$ such that the graph restricted to $S$ is symmetric, there are $(n-s)!$ elements in the equivalence graph $\hat{g}$. To illustrate, if $n=3$, there are exactly 16 network architectures for directed graphs.\footnote{See \url{http://mathinsight.org/image/three_node_motifs}. }

The network architecture is common knowledge. We assume that every agent observes, in addition to the network architecture, his position in the network. He may also observe other elements of the network, like the identity of his neighbors. Hence the information structure of  agent $i$  is a refinement of the information structure of the planner.  We let $E_i(g)$ the cell of agent $i$'s information partition at $g$.  We let ${\bf E}(g) = \{E_1(g),...,E_n(g)\}$ denote the set of possible information cells (or types) of the agents. To illustrate the information structures of the planner and the agents, consider the following network with three agents, where we assume that the agent's information structure is minimal (they only observe their location).

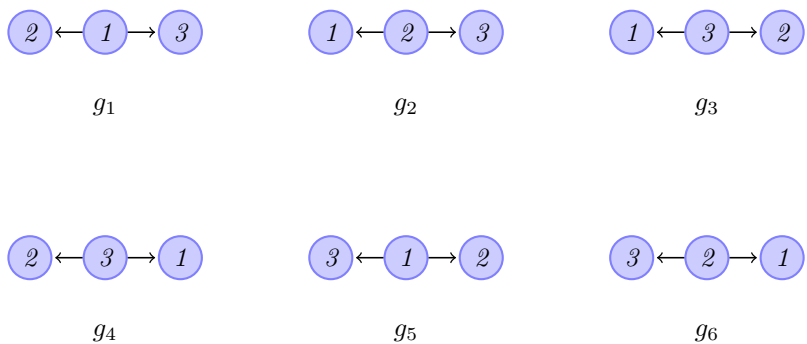
\begin{example}
\label{ex1}

\begin{equation*}
\begin{tikzpicture}[inner sep=1mm]
\tikzstyle{place}=[circle,draw=blue!50,fill=blue!20,thick,minimum size=2mm]
\tikzstyle{post}=[->,shorten >=1pt,semithick]
\tikzstyle{pre}=[<-,shorten <=1pt,semithick]
 \node at (0,4)[place](1){2}; \node at(1,4)[place](2){1} edge[post] node{}(1); \node at (2,4)[place](3){3} edge [pre] node{}(2);
 \node at (4,4)[place](4){1}; \node at(5,4)[place](5){2} edge[post] node{}(4); \node at (6,4)[place](6){3} edge [pre] node{}(5);
 \node at (8,4)[place](7){1}; \node at(9,4)[place](8){3} edge[post] node{}(7); \node at (10,4)[place](9){2} edge [pre] node{}(8);
 \node at (0,1)[place](10){2}; \node at(1,1)[place](11){3} edge[post] node{}(10); \node at (2,1)[place](12){1} edge [pre] node{}(11);
 \node at (4,1)[place](13){3}; \node at(5,1)[place](14){1} edge[post] node{}(13); \node at (6,1)[place](15){2} edge [pre] node{}(14);
 \node at (8,1)[place](16){3}; \node at(9,1)[place](17){2} edge[post] node{}(16); \node at (10,1)[place](18){1} edge [pre] node{}(17);
 \node at (1,3)(19){$g_1$}; \node at (5,3)(20){$g_2$}; \node at (9,3)(21){$g_3$}; \node at (1,0)(22){$g_4$}; \node at(5,0)(23){$g_5$}; \node at(9,0)(24){$g_6$};
 \end{tikzpicture}
\end{equation*}

The true network is network $g_1$. The planner is unable to identify agents in the network and hence will consider the equivalence class of $g_1$, $E(g_1)= \{g_1, g_2,g_3,g_4,g_5,g_6\}$. Agent $1$ knows that the network can either be $g_1$ or $g_5$ and hence identifies the cell in his information partition as $E_1(g_1)= \{g_1,g_5\}$. Similarly, agent $2$ identifies the cell as $E_2(g_1) = \{g_1,g_4\}$ and agent $3$ identifies the cell as $E_3(g_1) = \{g_1,g_2\}$. Notice that, by pooling the information of any pair of agents, the planner is able to exactly identify the network as $g_1$.

\end{example}

\subsection{The planner's mechanism}

We let the planner design a mechanism to extract information from the agents about their location in the network. For any network $g$, the planner learns the network architecture $\hat{g}$, and determines the list of possible types of the agents ${\bf E}(g) = \{E_1(g),...,E_n(g)\}$. Every agent $i$ sends to the planner an element in ${\bf E(g)}$, denoted $\hat{E}_i(g)$.  Given the vector of announcements ${\bf \hat{E}}(g) = (\hat{E}_1(g),..,\hat{E}_n(g))$, the planner selects a vector of decisions ${\bf x} = (x_1,..,x_n)$. 

We require the mechanism to be ex-post incentive compatible: for any network $g$, every agent should have an incentive to  report $E_i(g)$ rather than the cell of the partition corresponding to another agent. Formally

\begin{definition} A contract ${\bf x} = (x_1,...,x_n)$ is {\em ex-post incentive compatible} at a network architecture $\hat{g}$ if, for any $g \in \hat{g}$, for any  $i$, for any $E_j(g) \neq E_i(g) \in {\bf E(g)}$,
\[ U_i \big({\bf x}(E_i(g), E_{-i}(g)) \big) \geq U_i \big({\bf x}(E_j(g), E_{-i}(g))\big) .\]
\end{definition}

As we will see, correlation of agents' types makes it easy for the planner to construct incentive compatible contracts. We thus extend the definition of ex-post incentive compatibility to allow for joint deviations by coalitions of players. In the benchmark model, we assume that coalition $S$ must contain adjacent players. Formally

\begin{definition} A coalition $S$ contains adjacent players if and only if, for all $i,j \in S$, $g_{ij} + g_{ji} \geq 1$.
\end{definition}
The idea underlying this requirement is that, in order to engineer a joint deviation of reports, players in a coalition must communicate with each other, and hence be directly connected in the social network. In Section \ref{robust}, we will also discuss what happens when coalitions of nonadjacent players can misreport, and show that this may result in additional constraints on the contract.

We assume that if a coalition $S$ deviates, the identities of all members of $S$ are revealed to each other, so that agents in the deviating coalition can exactly evaluate their payoff following the deviation. (This assumption is reminiscent of the definition of the ``fine core" in Wilson (1978) \cite{wilson1978information}, where agents pool their information to engineer a deviation in an exchange economy with incomplete information). Absent this assumption, agents would be unable to identify the quantities consumed by their neighbors and hence the level of externalities they enjoy.  Notice that agents in $S$ will thus refine their information after the deviation is proposed. However, as the planner is unaware that the coalition $S$ forms and jointly misreports, she will be unable to exploit this fact in the design of the contract. 
\begin{definition} A contract ${\bf x} = (x_1,...,x_n)$ is {\em  ex-post group incentive compatible} at a network architecture $\hat{g}$ if, for any $g \in \hat{g}$, there does not exist a coalition $S$ containing adjacent players and a mapping $k(j)$ from $S$ to $N$  such that, for every agent $i$ in $S$,
\[ U_i \big({\bf x}((E _{k(j)}(g))_{j \in S}, (E_j(g))_{j \notin S} )\big) > U_i \big( {\bf x}((E_j(g))_{j \in S}, (E_j(g))_{j \notin S}) \big) .\]
\end{definition}

We will argue that this condition makes deviations very hard, because it requires all players to be strictly better off from the joint misreporting. A weaker notion, allowing for transfers across players in the deviating coalition $S$, will result in easier deviations (thereby making positive results on the optimal contract even stronger).

\begin{definition} A contract ${\bf x} = (x_1,...,x_n)$ is {\em  ex-post group incentive compatible with transfers} at a network architecture $\hat{g}$ if, for any $g \in \hat{g}$, there does not exist a coalition $S$ containing adjacent players and a mapping $k(j)$ from $S$ to $N$  such that,
\[ \sum_{i \in S} U_i({\bf x}(E _{k(j)} (g))_{j \in S}, (E_j(g))_{j \notin S}) > \sum_{i \in S} U_i({\bf x}(E_j(g))_{j \in S}, (E_j(g))_{j \notin S}) .\]
\end{definition}

\subsection{A menu of contracts}

In the benchmark case, we restrict the planner to offer a simple menu of contracts.  Fix one representative of the equivalence class $\hat{g}$, and let $l= 1,2,...,n$ denote the corresponding arbitrary labeling of the nodes. We call this labeling the  {\em locations} in the graph. Let $\ell: N \rightarrow N$ denote any permutation of the agents in $N$, which we interpret as an assignment of agents to locations.  The permutation corresponding to the specific network $g$ is denoted $\ell(g)$. Hence, for any agent $i$,  $\ell_i(g)$ denotes the true location of agent $i$ in the graph $g$.

The planner offers a menu of contracts $\{x_l\}$, corresponding to the decisions at each and every of the locations.  Every agent then selects a contract among the menu $\{x_l\}$. If all agents select different contracts, each agent obtains the contract he selected. Otherwise, if several agents select the same location, the planner punishes all agents and chooses an outcome ${\bf x} = {\bf 0}$. 

This simple menu of contracts only uses information about agents' locations, and not any additional information that the agents can have about the network (such as the identity of their neighbors). In Section \ref{robust}, we explore more complex contracts when agents have additional information on the network, and show that the planner will often be able to implement his first-best contract, exploiting the correlation between agents' types. However, we defend the menu of contracts as a very simple contract which is a realistic approximation of the contracts offered in reality. In particular, this contract  only relies on information on the network topology (the locations of the different nodes) and does not require any additional information, such as the identity of the agents. In addition, as we show below, if the agents only know their location in the network, and do not have any other information, the menu of contracts is equivalent to any other mechanism chosen by the planner.

\begin{proposition} \label{propequiv} Suppose that the agents only know their location in the network. If the mechanism $(x_1,..,x_l)$ is {\em incentive compatible} then every agent $i$ chooses the contract corresponding to his location, i.e.

\[ U_i \big( x(\ell_g)  \big) \geq U_i \big( x(k(i), \ell_{-i}(g)) \big), \mbox{ } \forall k(i) \neq \ell_i(j).\]

If the mechanism is group incentive compatible then  there does not exist a coalition $S$ of adjacent players and a mapping $k(j)$ from $S$ to $N$  such that, for every agent $i$ in $S$,
\[ U_i \big( x(k(j)_{j \in S}, \ell_j(g)_{j \notin S} )\big) \geq U_i \big( x(\ell_g) \big). \]

If the mechanism is group incentive compatible with transfers then  there does not exist a coalition $S$ of adjacent players and a mapping $k(j)$ from $S$ to $N$  such that,
\[ \sum_{i \in S} U_i \big( x(k(j)_{j \in S}, \ell_j(g)_{j \notin S} ) \big) > \sum_{i \in S} U_i(x(\ell_g)). \]
\end{proposition}

\noindent{\bf Proof of Proposition \ref{propequiv}:} We simply observe that $E_i(g) = \{g \in \hat{g} |, \ell_i =  \ell_i(g) \}$. There is a one-to-one relation between the messages sent by the agent in the general mechanism and the location of agents in the sorting game. In the general mechanism, any coalition $S$ of agents chooses a set of mapping assigning a location to each agent, $k(j)$ as in the sorting game. When all players announce $E_i(g)$, the planner associates one location to each agent, and hence all agents choose different contracts in the menu. If some agents choose to make the same announcement $E_i(g)$, they must obtain a low payoff in the mechanism, corresponding to the ${\bf 0}$ payoff in the menu of contracts.  \hfill $\square$

\subsection{Efficient incentive compatible contracts}

When agents do not coordinate their announcements, the planner can exploit correlation across types to implement an efficient contract. This result follows from the same intuition as in Cr\'{e}mer and McLean \cite{cremer1988full} who prove that efficient contracts can be implemented when agents' types are correlated. Whenever the planner receives inconsistent announcements, showing that one agent has misreported his type, she chooses to punish all agents. More specifically we prove the following result. 

\begin{proposition}  \label{optimalic} Any contract ${\bf x}$ where agents obtain positive payoffs is ex-post incentive compatible at any network architecture $\hat{g}$.
\end{proposition}

\noindent{\bf Proof of Proposition \ref{optimalic}:} Suppose that whenever the planner receives two announcements $E_i(g)$ corresponding to the same location $i$, she chooses to assign ${\bf 0}$. Then, for any agent $i$, 

\[ U_i (({\bf x}(E_i(g), E_{-i}(g))) = a  x_i - \frac{1}{2} x_i^2 + \alpha \sum_{j \in N} g_{ij} x_i x_j \geq 0 = U_i({\bf x}(E_j(g), E_{-i}(g))), \]

showing that the incentive compatibility constraint is always satisfied. \hfill $\square$

\bigskip

In particular, the optimal contract ${\bf x^*}$ is incentive compatible.

\subsection{Group incentive compatible contracts}

As individuals cannot gain by individually deviating from the truth, we now turn our attention to deviations by groups of agents. First notice, as in the proof of Proposition \ref{optimalic},  that if an agent $i$ in $S \subset N$ reports the information cell of an agent $j$ who does not belong to $S$, the planner will be able to detect that one of the agent lied, and punish all agents with a decision  ${\bf 0}$. Hence, if a deviation by a group of agents in $S$ is profitable, it must involve a report which is a permutation of the types  of agents inside $S$. Using the definition of the agent's utilities, this observation enables us to rewrite the constraints faced by the planner under group incentive compatibility.

\noindent A contract ${\bf x}= (x_1,..,x_n)$ is group incentive compatible if there does not exist a coalition $S$ of adjacent players and a permutation $\rho_S$ of agents in $S$ such that for all $i \in S$,
\[  a x_{\rho_S(i)} - \frac{1}{2} x_{\rho_S(i)}^2 + \alpha \sum_{j \in S}  g_{ij} x_{\rho_S(i )} g_{ij} x_{\rho_S(j)} + \alpha \sum_{j \notin S} g_{ij}  x_{\rho_S(i )} x_j > a x_i - \frac{1}{2} x_i^2 + \alpha \sum_{j \in N} g_{ij} x_i x_j. \]

We obtain the following important result:

\begin{proposition} The optimal contract ${\bf g^*}$ is group incentive compatible for any network architecture $\hat{g}$ if the size of the group $|S|$ is $2$, $3$ or $4$. \label{propair}
\end{proposition}

\noindent{\bf The proof of proposition \ref{propair}} is in  Appendix \ref{Appendix1}.

\bigskip

Proposition \ref{propair} shows that no coalition of size $2$, $3$ or $4$ can construct a group deviation, which makes all agents strictly better off. The logic of the Proposition is transparent in the case of deviations by pairs, when $|S|=2$.   The optimal contract ${\bf x^*}$ always assigns quantities which are higher than the optimal quantities of the agents, as the planner takes into account the positive externalities resulting from higher quantities. Hence, for a fixed set of quantities of the other players $k$, a player $i$ has no incentive to accept the quantity offered to player $j$ when $x_i^* < x_j^*$. When a pair $(i,j)$ of players exchanges quantities, the externalities they experience from each other remain identical, equal to $x_i^* x_j^*$. Hence there is no possibility for both players to benefit from the exchange as one of them must be accepting a higher quantity than the quantity he obtains by reporting truthfully his type. The same type of argument applies in a more complex way when $|S|=3$ or $|S|=4$  by considering different joint ways of misreporting, and focusing attention on those agents who tarde their quantity $x_i^*$ for a higher quantity. When $|S|$ is strictly greater than $4$, whether the optimal contract is group incentive for arbitrary graph architectures remains an open question. Proposition \ref{propair} suggests that asking for all agents to increase their payoff by a joint deviation might be too demanding. Hence it will often be important to study the weaker deviation concept, when transfers are allowed. \\

\noindent A contract ${\bf x}= (x_1,..,x_n)$ is group incentive compatible with transfers  if there does not exist a coalition $S$ of adjacent players and a permutation $\rho_S$ of agents in $S$ such that 
\[ \sum_{i \in S}  \big(a x_{\rho_S(i)} - \frac{1}{2} x_{\rho_S(i)}^2 + \alpha  \sum_{j \notin S} g_{ij} x_{\rho_S(i )} x_j \big) > \sum_{i \in S} \big( a x_i - \frac{1}{2} x_i^2  + \alpha  \sum_{j \notin S} g_{ij} x_i x_j \big)\]

To interpret the last condition, notice  that by definition $\sum_{i \in S} \sum_{j \in S} g_{ij} x_{\rho_S(i )} x_{\rho_S(j)} = \sum_{i \in S} \sum_{j \in S} g_{ij} x_i x_j$. In addition, as agents are homogeneous, $\sum_{i \in S} (a x_{\rho_S(i)} - \frac{1}{2} x_{\rho_S(i)}^2 ) = \sum_{i \in S} (a x_i - \frac{1}{2} x_i^2$), and the condition for group incentive compatible contracts with transfers reduces to

\begin{equation}
\label{groupict}
\sum_{i \in S} \sum_{j \notin S} g_{ij} x_{\rho_S(i)} x_j \leq \sum_{i \in S}  \sum_{j \notin S} g_{ij} x_i x_j \forall S, \mbox{ } \forall \rho_S
\end{equation}

\section{Undirected Networks}

In this Section, we consider undirected network architectures,  where agents' influence is reciprocal, and the adjacency matrix ${\bf G}$  is symmetric. This corresponds for example to team production with synergy across workers, or consumption externalities for agents using the same software, same standard or same communication device. Our main Proposition shows that, in that case, the optimal contract is group incentive compatible with transfers.

\begin{proposition} \label{undirected}  If the network $g$ is undirected, then the optimal contract ${\bf x^*}$ is group incentive compatible with transfers.
\end{proposition}

\noindent{\bf Proof:} We first show that there does not exist a pair of connected agents $(i,j)$ who can benefit from a deviation. Suppose that this were the case, then we would have
\[ x^*_j \sum_{k \neq j \in N} g_{ik} x^*_k + x^*_i \sum_{k \neq i \in N} g_{jk} x^*_k > x^*_i \sum_{k \neq j \in N} g_{ik} x^*_k + x^*_j \sum_{k  \neq i \in N} g_{jk} x^*_k, \]

\noindent As $g_{ij} = g_{ji} =1$, we can add $2 x^*_i x^*_j$ on both sides of the inequality, to obtain
\[ x^*_j \sum_{k \in N} g_{ik} x^*_k + x^*_i \sum_{k \in N} g_{jk} x^*_k > x^*_i \sum_{k \in N} g_{ik} x^*_k + x^*_j \sum_{k \in N} g_{jk} x^*_k, \]

\noindent and collecting terms

\begin{equation}
\label{eqpair}
 (x^*_j - x^*_i) (\sum_{k \in N} g_{ik} x^*_k - \sum_{k \in N} g_{jk} x^*_k) > 0. 
\end{equation}

\noindent Now recall that ${\bf x^*} = ({\bf I} - \alpha {\bf G})^{-1} {\bf 1}$, the Katz-Bonacich centrality measure vector. By construction, the Katz-Bonacich centrality of an agent $i$ can be recursively expressed as a function of the Katz-Bonacich centrality of agent $i$'s neighbors:

\[ b_i ({\bf G}, \alpha) = 1 + 2 \alpha \sum_{k \in N} g_{ik} b_k({\bf G}, \alpha) .\]

But then, if $b_j({\bf G}, \alpha) > b_i({\bf G}, \alpha)$, we must have $\sum_{k \in N} g_{jk} b_k({\bf G}, \alpha) > \sum_{k \in N} g_{ik} b_k({\bf G}, \alpha), $ contradicting equation (\ref{eqpair}).

In the second step of the proof, we show that there cannot exist a coalition of adjacent players of size $k > 2$ which has a profitable deviation.  Suppose that it were the case and let $\rho$ denote the permutation of players in $S$ which results in a profitable deviation. We must then have

\[ \sum_{i \in S} \sum_{j \notin S} g_{ij} x_{\rho_S(i)} x_j  >  \sum_{i \in S}  \sum_{j \notin S} g_{ij} x_i x_j.\]

Because $g_{ij}=1$ for all $i,j \in S$, we can add to the left and right sides of the inequality $2 \sum_{i,j \in S} x_i x_j = 2 \sum_{i,j \in S} x_{\rho_S(i)} x_{\rho_S(j)}$ to obtain

\begin{equation}
\label{eqmult1} 
 \sum_{i \in S} \sum_{j \neq i } g_{ij} x_{\rho_S(i)} x_j > \sum_{i \in S} \sum_{j \neq i } g_{ij} x_i x_j.
 \end{equation}

Next order the vector of ${\bf x}$ so that $x_i \leq x_{i+1}$ for all $i \in S$ and let all indices outside $S$ be larger than the indices in $S$. Then we can rewrite equation (\ref{eqmult1}) as

\begin{equation}
\label{eqmult2}
(x_{\rho_S(1)} - x_1) \sum_j g_{1j} x_j + (x_{\rho_S(2)} - x_2) \sum_j g_{2j} x_j +.....+ (x_{\rho_S(s)} - x_s) \sum_j g_{sj} x_j > 0. 
\end{equation}

Next consider the set of agents in $S$ such that $\rho(i) \neq i$, discard all other agents in $S$ and reorder if needed the indices so that all remaining agents in $S$ have lower indices. Let $t$ be the size of the remaining set of indices. For any $j \in \{1,..,t\}$ define the following two sets of indices:
\begin{eqnarray*}
A_j& = & \{i| \rho(i) \geq j+1 > j \geq i \}, \\
B_j = & = & \{i| i \geq j+1 > j \geq \rho(i)\}.
\end{eqnarray*}

Clearly $A_j \cap B_j = \emptyset$. We now show by induction that $|A_j| = |B_j|$  and that $\max \{i|i  \in A_j\} < \min \{i|i \in B_j\} $ for all $j$. 

Consider the initial step at $j=1$. Then clearly $\rho(1)>1$ and we have $A_1 = \{1\}$. Let $k$ be the unique antecedent of $1$, i.e. $\rho(k)=1$. Then $B_1 = \{k\}$ and hence $|A_1|=|B_1|=1$ and $k>1$.

Next assume that $|A_{j-1}|=|B_{j-1}|$ and $\max \{i|i  \in A_{j-1}\} < \min \{i|i \in B_{j-1}\}$. Consider the sets $A_j$ and $B_j$. By construction we have

\begin{eqnarray*}
A_j & = & A_{j -1}\setminus \{i| \rho(i)=j > j-1 \geq i\} \cup \{i| \rho(i) \geq j+1>j=i\} \\
B_j & = & B_{j -1}\setminus \{i|i =j > j-1 \geq \rho(i)\} \cup \{i| i \geq j+1 > j = \rho(i) \}
\end{eqnarray*}

Now let $k$ be the antecedent of $j$ in the permutation $\rho$, i.e. $\rho(k)=j$. We consider the four following cases:

\medskip

\noindent{\em Case 1. $\rho(j) > j, j>k$} In that case $A_j = A_{j -1}\setminus \{k\} \cup \{j\}, B_j = B_{j-1}$ and hence $|A_j| = |B_j|.$ Furthermore, $j > i$ for all $i \in A_{j-1}$ so $j = \max \{i|i \in A_j\}$. Finally, $i \geq j$ for all $j \in B_{j-1}$ and as $j \notin B_j$, $\min \{i|i \in B_j \} > j = \max \{i|i \in A_j\}$.

\medskip

\noindent {\em Case 2. $\rho(j)>j, k> j$} In that case $A_j = A_{j-1} \cup \{j\}, B_j = B_{j-1} \cup \{k\}$ and hence $|A_j| = |B_j|.$ As $k>j$, if $\max \{i|i \in A_{j-1} \} < \min \{i| i \in B_{j-1} \}$, we must also have  $\min \{i|i \in B_j \} > j = \max \{i|i \in A_j\}$.

\medskip

\noindent {\em Case 3. $\rho(j)<j, j> k$} In that case $A_j = A_{j -1}\setminus \{k\}, B_j = B_{j-1} \setminus \{j\}$ and hence $|A_j| = |B_j|.$ Furthermore, as $A_j \subset A_{j-1}$, $\max \{i| i \in A_j\} \leq \max \{i| i \in A_{j-1}\}$ and as $B_j \subset B_{j-1}$, then $\min \{i|i \in B_j\} \geq \min \{i|i \in B_{j-1} \}$. Hence $\min \{i|i \in B_j\} \geq \max \{i| i \in A_j\} $.

 \medskip
 
\noindent{\em Case 4. $\rho(j)<j, k>j$} In that case $A_j = A_{j-1}$ and $B_j= B_{j-1} \setminus \{j\} \cup \{k\}$ and hence $|A_j| = |B_j|$. In addition as $k > j$, $\min \{i|i \in B_j\} \geq \min  \{i|i \in B_{j-1}\} > \max \{i|i \in A_{j-1}  \} = \max \{i|i \in A_j\}$.

\medskip

Now rewrite equation (\ref{eqmult2}) as

\begin{equation}
\label{eqmult3}
 \sum_{j=1}^{t-1} (x_{j+1} -x_j) ( \sum_{i \in A_j} \sum_k g_ik x_k  - \sum_{i \in B_j} \sum_k g_{ik} x_k ) > 0. 
 \end{equation}

Now because $A_j| = |B_j|$, we can pick for any $i \in A_j$ a corresponding index $l$ in $B_j$ and as $l > i$, we also have $x_l > x_i$. But by equation (\ref{eqpair}), if $(i,l)$ does not have a profitable deviation and $x_l > x_i$ we must have

\[ \sum_k g_{lk} x_k \geq \sum_k g_{ik} x_k.\]

But this implies that for all $j$

\[ ( \sum_{i \in A_j} \sum_k g_ik x_k  - \sum_{i \in B_j} \sum_k g_{ik} x_k ) \leq 0, \]

\noindent contradicting equation (\ref{eqmult3}). This contradiction completes the proof of the Proposition.

Proposition \ref{undirected} shows that when influence is reciprocal, coalitions of adjacent agents cannot benefit from jointly misreporting their locations when the planner proposes the optimal contract. This result is due to the fact that in an undirected network agents with higher levels in the first-best contract are agents who receive and produce the larger externalities. Hence, constructing a group deviation against the first-best contract becomes impossible, because agents with higher levels are also agents with higher externalities. This is a very strong result because the definition of coalitional deviations is very permissive -- we allow for any transfer across coalition members. The only restriction we place on the deviating coalition is that it should contain adjacent agents. As the proof demonstrates, if we only consider pairs of agents swapping their announcements, this condition is not needed: the optimal contract is robust to any deviation by pairs of players, whether they are adjacent or not. However, for larger coalitions, the restriction that only adjacent players can form a deviating coalition is meaningful, as we will illustrate in section \ref{robust}

\section{Hierarchical Networks}

We next consider very asymmetric structures.  Agents are organized in a hierarchy. Agents at higher levels of the hierarchy influence agents below them, without being influenced by them. For any pair of agents, externalities only flow in one direction, from agents at higher tiers to agents at lower tiers. Formally, we partition the set of agents into $M$ tiers of a hierarchy,  $A_1,..,A_M$ with $M \geq 2$ such that $g_{ij}=1$ if and only if  $i \in A_m, j \in A_q$ and $ q <m$. \footnote{We later consider an alternative formulation where $q \leq m$, i.e. agents in the same level of the hierarchy influence each other.} 

\begin{proposition} In a hierarchical network, the optimal contract ${\bf x^*}$ is group incentive compatible. \label{hiergic}
\end{proposition}

\noindent{\bf Proof:}  Suppose that there exists a coalition $S$ of adjacent agents and a permutation $\rho$ such that every agent $i$ in $S$ has a higher payoff by exchanging locations according to $\rho$ at the optimal contract ${\bf x^*}$. There must be a set of agents $S^-$ for whom $x_{\rho(i)}^* > x_i^*$. Among those agents, pick an agent $i$ belonging to the highest level of the hierarchy so that $g_{ij}=0$ for any $j \in S^{-}$. For that agent, $(x^*_{\rho(j)}-x_j) <0 $ for any $j \in S,$ such that $g_{ij}=1$. Hence, as $(x^*_{\rho(i)}-x^*_i) > 0$, 

\begin{eqnarray*}
\Delta U_i & = &  \alpha [- \sum_{j \notin S} (x^*_{\rho(i)}-x^*_i) g_{ji} x_j^* - \frac{(x^*_{\rho(i)}-x^*_i)^2}{2} + \sum_{j \in S} g_{ij} x^*_{\rho(i)}(x^*_{\rho(j)}-x_j) - \sum_{j \in S} g_{ji} x^*_j (x^*_{\rho(i)}-x_i^*)].\\
& < & 0
\end{eqnarray*}

\noindent contradicting the fact that agent $i$ has an incentive to deviate. \hfill $\square$

\bigskip

Proposition \ref{hiergic} shows that groups of players, even if they are larger than three, cannot organize exchanges of locations which make all the agents strictly better off. Among agents who are assigned higher quantities than the first-best quantity, the agent in the highest tier of the hierarchy only enjoys externalities from agents who exchange their quantities for lower quantities. Hence the agent can surely not increase his utility by receiving a higher quantity. 

\subsection{Single root} 

While Proposition \ref{hiergic} shows that the optimal contract is immune to group deviations in hierarchical networks, we will now see that, contrary to the case of undirected networks, the sum of utilities of agents in a deviating coalition can increase. In order to analyze group incentive compatible contracts with transfers, we first consider hierarchical networks with a single root.

\begin{proposition} \label{hiersingle} Suppose that the hierarchy has a single root, $|A_1|=1$ and that all agents are connected to the root and that the hierarchy contains more than two tiers. Then agents at tiers $m=3,..,M$ must receive the same quantity as the root agent in the optimal contract satisfying group incentive compatibility with transfers.
\end{proposition}

\noindent{\bf Proof:}   Consider a two-player deviation with transfers between any agent $i$ in $A_m$, $m > 2$ and the root agent $j$  in $A_1$. We must have

\[ \sum_{k \neq j} g_{ik} x_j  x_k + \sum_{k \neq i} x_i x_k \leq  \sum_{k \neq j} g_{ik} x_i  x_k + \sum_{k \neq i} g_{jk} x_j x_k, \]

\noindent Next, as $g_{jk}=0$ for all $k$, this condition amounts to

\[  \sum_{k \neq j} g_{ik} x_j  x_k \leq \sum_{k \neq j} g_{ik} x_i  x_k, \]

\noindent or $x_j \leq x_i$. Notice that this condition must hold for any pair $(i,j)$ where $j$ is the root agent and  $i \in A_k$, $k > 2$. 

Suppose  that the planner selects a contract ${\bf x}$ such that $x_j < x_i$ for some $ i \in A_m$ for $m > 2$. Consider the marginal effect of increasing $x_j$ by $\epsilon$ and reducing $x_i$ by $\epsilon$ on the sum of payoffs:

\begin{eqnarray*}
  \sum \Delta U & = & - \frac{(x_i- \epsilon)^2 - x_i^2}{2} - \frac{(x_j+ \epsilon)^2 - x_j^2}{2} - \alpha \epsilon \sum_k (g_{ik} + g_{ki}) x_k + \alpha  \epsilon \sum_k (g_{jk} + g_{kj}) x_k. \\
 & = & \epsilon (x_i - x_j )- \epsilon ^2 - \alpha \epsilon \sum_k (g_{ik} + g_{ki}) x_k + \alpha  \epsilon \sum_k (g_{jk} + g_{kj}) x_k.
 \end{eqnarray*}
 
 Now recall that in a hierarchical network, $g_{ik}+g_{jk} \leq 1$. Because player $j$ is connected to all agents,  $g_{jk} + g_{kj} = 1$ for all $k \neq j$. Player $i$ is at best connected to all agents  so $g_{ik} + g_{ki} \leq 1 $ for all $k \neq i$. Now
 
 \begin{eqnarray*}
  \sum_k (g_{jk} + g_{kj}) x_k - \sum_k (g_{ik} + g_{ki}) x_k  & = &  (x_i - x_j) + \sum_{k \neq i,j} (1 - (g_{ik} + g_{ki})) x_k, \\
  & \geq & (x_i - x_j) \\
  & > & 0
  \end{eqnarray*}
  
  \noindent so that
  
  \[ \sum \Delta U > 0. \]
  
  \noindent for small enough $\epsilon$. This shows that the optimal contract satisfying group incentive compatible with transfers must assign the same quantity to $x_i$ and $x_j$. \hfill $\square$.
  
  \bigskip
  
 Proposition \ref{hiersingle}  shows that agents in tiers $m=3,..,M$ must receive the same quantity as the root agent. Suppose by contradiction that  the root agent receives a higher quantity. We claim that she has an incentive to exchange her location with any other agents at a tier $m \geq 3$, as those agents receive positive externalities form agents at tier $2$, in addition to the positive externalities from the root agent, whereas the root agent does not receive any positive externality. Hence a contract which satisfies group incentive compatibility with transfers must assign to the root agent a quantity which is lower or equal than the quantity of agents at tiers $m \geq 3$. As the sum of utilities increases when the quantity assigned to the root agent increases, this implies that the best contract must assign uniform quantities to agents in tiers $m=1,3,..,M$. 
 
 As long as not all agents are connected to each other (in which case they all receive the same quantity at the optimal contract ${\bf x^*}$), some agents must receive different quantities at the optimal contract ${\bf x^*}$. This shows that {\em the optimal contract is not group incentive compatible with contracts}. The incentive constraint binds and forces the planner to equalize quantities offered agents at different tiers of the hierarchy.
 
 Interestingly, Proposition \ref{hiersingle} does not restrict the quantities offered at tier $m=2$. Agents at tier $m=2$ have no incentive to exchange their location with the root agent, as the sum of utilities remains exactly the same after the exchange. Whether agents at tier $2$ have an incentive to exchange locations with agents at lower tiers depends on the precise structure of the network, and cannot be ascertained in general. Hence, in general, the choice of quantities to agents in the second tier of the network remains unrestricted.
 
 Finally, we note that, because quantities at the second level of the hierarchy are unrestricted, Proposition \ref{hiersingle} does not hold when $M=2$. In that case, the group incentive constraint with transfers is not binding, and the optimal contract ${\bf x^*}$ is immune to group deviations with transfers. 
  
 We next consider a situation where either the first level of the hierarchy contains multiple roots, or agents are not all connected to the single root. 
 
 \subsection{Nested neighborhoods}
 
We first analyze a hierarchy with multiple roots, where neighborhoods are nested, so that agents at lower tiers of the hierarchy are influenced by the same agents as agents at higher tiers of the hierarchy. 

Formally, for any three tiers $m,p,q$ with $m < p< q$, we have

\[ \forall i, \in A_q, \forall j \in A_p, \mbox{ such that } g_{ij}=1, \{k| g_{ik} =1 \} \cup A_m = \{k| g_{jk} =1 \} \cup A_m. \]

This definition covers the case where agents at tier $q$ are influenced by {\em all agents at higher tiers of the hierarchy}. But it also covers cases where agents are only influenced by a subset of agents at higher tiers of the hierarchy, as long as the influence relation is equal to its transitive closure: namely, for $i \in A_q, j \in A_p, k \in A_m$, $g_{ik}=1$ if and only if $g_{ij}=1$ and $g_{jk}=1$. Figure \ref{fignested} illustrates a hierarchy with nested neighborhoods and three tiers.

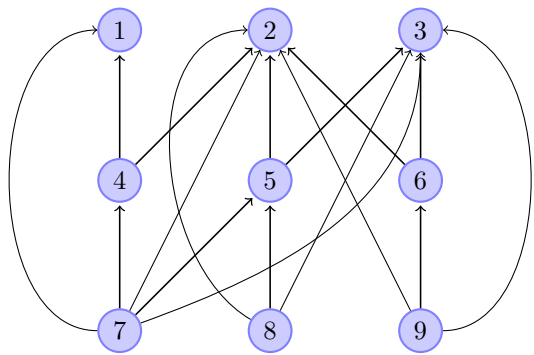
\begin{equation*}
\begin{tikzpicture}[inner sep=1mm]
\tikzstyle{place}=[circle,draw=blue!50,fill=blue!20,thick,minimum size=2mm]
\tikzstyle{post}=[->,shorten >=1pt,semithick]
\tikzstyle{pre}=[<-,shorten <=1pt,semithick]
 \node at (0,2)[place](1){1}; \node at (2,2)[place](2){2}; \node at (4,2)[place](3){3};
 \node at(0,0)[place](4){4} edge[post] node{}(1) edge[post] node{}(2); \node at (2,0)[place](5){5} edge[post] node{}(2) edge[post] node{}(3); \node at (4,0)[place](6){6} edge[post] node{}(2) edge[post] node{}(3);
 \node at(0,-2)[place](7){7} edge[post] node{}(4) edge[post] node{}(5); \node at (2,-2)[place](8){8} edge[post] node{}(5); \node at (4,-2)[place](9){9} edge [post] node{}(6);
 \draw[->] (7) to [out=180,in=180] (1); \draw[->] (7) to (2); \draw[->](7) to [out=20,in=270] (3);
 \draw[->](8) to [out=150,in=180] (2); \draw[->](8) to (3);
 \draw[->](9) to (2); \draw[->](9) to [out=0,in=0] (3);
 \label{fignested}
 \end{tikzpicture}
\end{equation*}

\begin{proposition} \label{hiernested} Consider a hierarchy with nested neighborhoods  such that all agents are connected to more than one root, and $|A_m| \leq |A_{m+1}|$ for all $m=1,..,M-1$. Then all agents receive the same quantity in the optimal contract satisfying group incentive compatibility with transfers.
\end{proposition}

\noindent{\bf Proof:} Let $i$ be  any agent at a tier $A_m$, $m \geq 2$ and $j$ be a root agent. For the contract to be immune to a deviation by the pair $(i,j)$, we must have

\[  \sum_{k \neq j} g_{ik} x_j  x_k \leq \sum_{k \neq j} g_{ik} x_i  x_k, \]

\noindent Because every agent is connected to at least two root agents, $\sum_{k \neq j} g_{ik} \neq 0$, so that we must have

\[ x_j \leq x_i.\]

Now, suppose by contradiction that $x_j < x_i$ and compute the effect of a small decrease in $x_i$ couple with a small increase in $x_j$ on the sum of utilities:

\[ \sum \Delta U = \epsilon (x_i - x_j )- \epsilon ^2 - \alpha \epsilon \sum_k (g_{ik} + g_{ki}) x_k + \alpha  \epsilon \sum_k (g_{jk} + g_{kj}) x_k.\]

To complete the proof of the Proposition, we need to show $\sum_k  g_{kj} x_k \geq \sum_k (g_{ik} + g_{ki}) x_k $. Consider first agents $k \in A_q$ with $q>m$ so that $g_{ik}=0$. By the definition of nested neighborhoods, if $g_{ki}=1$ then $g_{kj}=1$, so

\[ \sum_{k \in A_q} g_{kj}x_k \geq \sum_{k \in A_q} g_{ki} x_i.\]

\noindent Consider next agents $k \in A_q$, $1<q<m$. Suppose that $g_{ik}=1$ and $g_{kj}=0$. Then, by the definition of nested neighborhoods we must have $g_{ij}=0$, contradicting the fact that $i$ is connected to $j$. So we must have $g_{kj}=1$ and hence again

\[ \sum_{k \in A_q} g_{kj}x_k \geq \sum_{k \in A_q} g_{ki} x_i.\]

\noindent We finally consider $k \in A_1, k \neq j$ and $k \in A_m \neq i$.  Recall that if $k \in A_1$ and $k' \in A_m$, $x_k \leq x_k'$.  Furthermore, by assumption $|A_m| \geq |A_1$, so that
\[ \sum_{k' \in A_m} g_{k'j} x_{k'} \leq \sum_{k \in A_1} g_{ik} x_k, \]

\noindent completing the proof of the Proposition. \hfill $\square$

\bigskip

Proposition \ref{hiernested} displays a family of hierarchies with multiple roots for which the group incentive compatibility constraint forces the planner to select {\em uniform} quantities in the network. The planner is unable to target agents at different locations, as agents always have an incentive to exchange their locations with one of the root agents. This may result in vary large efficiency losses for the planner, as the optimal contract clearly discriminates among agents, offering higher quantities to agents at higher tiers in the hierarchy. Observe that Proposition \ref{hiernested} relies on the fact that the number of agents is nondecreasing at lower levels of the hierarchy. As the following example shows, if this condition fails, the optimal contract may very well be group incentive compatible with transfers.

\begin{example}
\label{exdecreasing}
Suppose that  $n=4$, agents $1,2,3$ belong to the first level of the hierarchy and  a single agent $4$ to the second level $A_2$, i.e. $g_{41}=g_{42}=g_{43}=1$ and $g_{ij}=0$ for all other $ij$. Then in the first-best contract we have

\begin{eqnarray*}
x_1^* & = & a +  \alpha x_4^*, \\
x_4^* & = & a + 3 \alpha x_1^*, 
\end{eqnarray*}

\noindent so that 

\[ x_1^* = x_2^* = x_3^* = \frac{a(1+\alpha)}{1-3 \alpha^2} < x^*_4 = \frac{a(1+ 3 \alpha)}{1- 3 \alpha^2}. \]

The optimal contract assigns a higher quantity to the agent at the lower level $A_2$, so that the group incentive compatibility with transfers constraint holds.
\end{example}

\subsection{Regular Oriented Trees}

We next consider hierarchies with a single root, but where agents are not all connected to the root. Instead, the hierarchy is a tree oriented towards the root, so that any node $i$ has a {\em single} neighbor in the directed network, located at the tier immediately above him. We let $\sigma_i$ denote the set of immediate predecessors of agent $i$ in the oriented tree.  We suppose that at any tier $A_m$ of the hierarchy, $|\sigma_i| = |\sigma_j|$ for all $i,j \in A_m$. Hence the oriented tree is {\em regular}: there are the same number of predecessor nodes for any node at the same tier of the tree.  Figure \ref{figortree} illustrates a regular tree oriented towards the root with $n=7$ agents. In this example $\sigma_1 =\{2,3,4\}, \sigma_2 = \{5\}, \sigma_3 = \{6\}, \sigma_4=\{7\}, \sigma_5 =  \sigma_6 = \sigma_7 =\emptyset$.

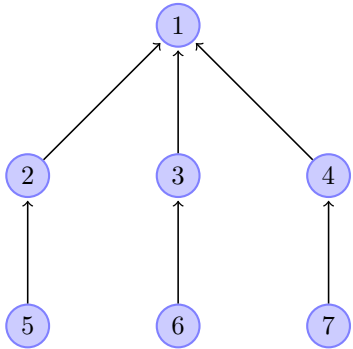
\begin{equation*}
\begin{tikzpicture}[inner sep=1mm]
\tikzstyle{place}=[circle,draw=blue!50,fill=blue!20,thick,minimum size=2mm]
\tikzstyle{post}=[->,shorten >=1pt,semithick]
\tikzstyle{pre}=[<-,shorten <=1pt,semithick]
 \node at (2,2)[place](1){1}; 
 \node at(0,0)[place](2){2} edge[post] node{}(1); \node at (2,0)[place](3){3} edge [post] node{}(1); \node at (4,0)[place](4){4} edge [post] node{}(1);
 \node at(0,-2)[place](5){5} edge[post] node{}(2); \node at (2,-2)[place](6){6} edge [post] node{}(3); \node at (4,-2)[place](7){7} edge [post] node{}(4);
 \label{figortree}
 \end{tikzpicture}
\end{equation*}

The following Proposition shows that, when the number of predecessors is higher for higher tiers in the oriented tree, the optimal contract is immune to group deviations with transfers.

\begin{proposition} \label{ortree} Consider a regular oriented tree such that,  $|\sigma_i| < |\sigma_j|$ if $i \in A_m, j \in A_q$ and $m < q$. Then the optimal contract ${\bf x^*}$ is group incentive compatible with transfers.
\end{proposition}

\noindent{\bf Proof:} Because every agent has a single neighbor, the only adjacent coalitions are pairs of agents $(i,j)$ such that $g_{ij}=1$. We show that the optimal contract is immune to deviations by pairs of players with transfers. First note that if $j \in A_1, i \in A_2$, then, as there is no $k \neq j$ such that $g_{ik}=1$ and by definition  $g_{jk}=0$ for all $k$. Hence, any pair of quantities is group incentive compatible with transfers. 

Next pick two agents $j,i$ in $A_m, A_{m+1}$ with $m \geq 2$. Notice that there is no agent different from $j$ such that $g_{ik}=1$ whereas $g_{jk}=1$ for some agent $k \in A_{m-1}$. Hence the condition

\[ \sum_{k \neq j} g_{ik} x_j  x_k + \sum_{k \neq i} x_i x_k \leq  \sum_{k \neq j} g_{ik} x_i  x_k + \sum_{k \neq i} g_{jk} x_j x_k, \]

\noindent is equivalent to

 \begin{equation} \label{eqortree}
 x_i \leq x_j \mbox{  for all }  i,j   \mbox{  such  that  }  g_{ij}=1, j \neq A_1.
 \end{equation}
 
 \noindent Hence any contract for which $x_i \leq x_j$ whenever $g_{ij}=1$ is group incentive compatible with transfers. 
 
 Next, we show that  the optimal contract ${\bf x^*}$ satisfies condition (\ref{eqortree}). The proof is by induction on the tier of the hierarchy. Notice that because the tree is regular, $x^*_i = x^*_j$ for any $i,j$ at the same tier of the hierarchy. Hence, we let $x^{*m}$ denote the common quantity at tier $m$ of the hierarchy.
 
 Consider $(i,j)$ such that $g_{ij}=1$ and $i \in A_M, j \in A_{M -1}$. Then $|\sigma_i|= \emptyset$. Let $l$ be the successor of $j$ in the oriented tree. The optimal contract satisfies
 
 \begin{eqnarray*}
 x_i^* & = & a + \alpha x_j^*, \\
 x_j^* & = & a + \alpha x_i^* + \alpha \sum_{k | g_{kj}=1} x_k^* + \alpha x_l^*
 \end{eqnarray*}
 
 \noindent so that
 
\[  (x_i^* - x_j^*) (1-\alpha) = - \alpha \sum_{k | g_{kj}=1} x_k^* + \alpha x_l^* <0.\]

This shows that $x^{*M} < x^{*M-1}$.

Next suppose that $x^{*q} < x^{*q-1}$ for any $q > m$ and consider a pair $(i,j)$ with $g_{ij}=1$,  $i \in A_m, j \in A_{m-1}$. Let $l$ be the successor of $j$. The optimal contract satisfies

\begin{eqnarray*}
x^*_i & = & a + \alpha x^*_j + \alpha \sum_{k | g_{ki}=1} x^*_k, \\
x^*_j & = & a+ \alpha x^*_i + \alpha \sum_{k \neq i, g_{kj} = 1} x^*_k + \alpha x^*_l.
\end{eqnarray*}

\noindent Hence

\[ (x_i^* - x_j^*) (1-\alpha) = - \alpha (\sum_{k \neq i, g_{kj} = 1} x^*_k - \sum_{k|g_{ki}=1} x^*_k) - \alpha x^*_l.\]

Now, by assumption $|\sigma_j| \geq |\sigma_i|+1$, so $| \{k \neq i, g_{kj}=1\}| \geq | \{k| g_{ik} = 1\}|$. In addition, by the induction hypothesis, $x^*_k > x^*_l$ for any $k,l$ such that $g_{ki} = g_{lj}=1$ as $k \in A_{m+1}, l \in A_m$ and by the induction hypothesis $x^{*m} > x^{*m+1}$. This shows that $x_i^* < x_j^*$, completing the proof of the Proposition. \hfill $\square$

\bigskip

Proposition \ref{ortree} shows that in a regular oriented tree, the optimal contract can be sustained by the planner. When every agent is influenced by a single predecessor, the contract is immune to exchange by two adjacent agents if and only if the predecessor has a greater quantity than the successor. The optimal contract ${\bf x^*}$ assigns to every agent a quantity which is proportional to her Katz-Bonacich centrality measure in the undirected tree. When the tree is regular, the Katz-Bonacich centrality measure will be higher at higher tiers  whenever the number of agents connected to another agent is lower at lower tiers of the hierarchy.

Notice that the assumption that the number of agents connected to another agent is lower at lower levels of the hierarchy is crucial for the proof of Proposition \ref{ortree}. Consider for example a line of five  agents, $n=5$, $g_{12} = g_{23} = g_{34} = g_{45} = 1$, $g_{ij} = 0$ for all other pairs $(i,j)$. It is easy to check that the Katz-Bonacich centrality measure is highest at the center of the line, and decreases towards the edges, so that $x^*_3 > x^*_2=x^*_4 > x^*_1 = x^*_5$. But then the condition $x^*_4 \geq x^*_3$ is violated, and the optimal contract is not immune to a deviation by the two players, $3$ and $4$.

\subsection{Connected agents at the same tier of the hierarchy}

Next consider an alternative formulation, where agents are influenced by other agents at the same level of the hierarchy, i. e $g_{ij}=1$ if $i, j \in A_m$. In that case, we do not know whether Proposition \ref{hiergic} holds, as the argument showing that the optimal contract is group incentive compatible relies on the fact that the agent at the highest level of the hierarchy in $S^-$ does not have any connection to other agents in $S^-$. Proposition \ref{hiersingle} may also be violated, as the first-best contract can assign a higher quantity to agents at lower levels of the hierarchy. For example, suppose that $n=5$, $A_1 = \{1\}$ , $A_2 = \{2\}, A_3 = \{3,4,5\}$. We have $g_{21}= g_{31}=g_{32}=g_{34}=g_{35} =g_{41} = g_{42}=g_{43}=g_{45} = g_{51}=g_{52}=g_{53}=g_{54}=1$ and $g_{ij}=0$ otherwise. The first-best contract solves

\begin{eqnarray*}
x_1^* & = & a +  \alpha x^*_2 + 3 \alpha x^*_3, \\
x_2^* & = & a +  \alpha x^*_1 + 3  \alpha x^*_3, \\
x^*_3 & = & a + \alpha x^*_1 + \alpha x^*_2 + 4 \alpha x^*_3
\end{eqnarray*}

\noindent so that

\[ x^*_1 =x^*_2= \frac{a (1- \alpha)}{1- 5 \alpha - 2 \alpha^2} < x^*_3 = x^*_4 = x^*_5 = \frac{a(1+\alpha)}{1- 5 \alpha - 2 \alpha^2}, \]

\noindent and the first-best contract satisfies the group incentive compatibility with transfers condition.

\section{Robustness and Extensions} \label{robust}

\subsection{Nonadjacent agents} When the deviating coalitions comprise any set of players (not necessarily adjacent), the first-best contract may fail to be immune to group deviations with transfers. We illustrate this fact by looking at the two families of network architectures for which the first-best contract has been shown to be immune to deviations by coalitions of adjacent agents: undirected networks and regular oriented trees.

\subsubsection{Undirected networks}

\begin{example}
\label{exdevund}
Consider 8 agents organized in two disjoint stars as in the following picture

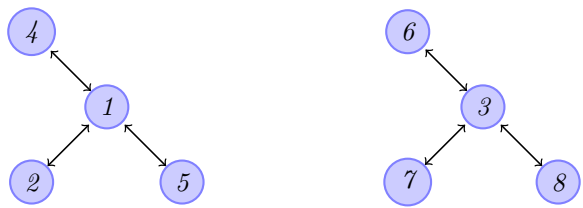
\begin{equation*}
\begin{tikzpicture}[inner sep=1mm]
\tikzstyle{place}=[circle,draw=blue!50,fill=blue!20,thick,minimum size=2mm]
\tikzstyle{arrow}=[<->,shorten >=1pt,semithick]
 \node at (0,2)[place](1){4}; \node at(1,1)[place](2){1} edge[arrow] node{}(1); \node at (0,0)[place](3){2} edge [arrow] node{}(2); \node at (2,0)[place](4){5} edge [arrow] node{}(2);
 \node at (5,2)[place](5){6}; \node at(6,1)[place](6){3} edge[arrow] node{}(5); \node at (5,0)[place](7){7} edge [arrow] node{}(6); \node at (7,0)[place](8){8} edge [arrow] node{}(6);
 \end{tikzpicture}
\end{equation*}
\end{example}

In a star with three peripheral nodes, the Katz-Bonacich centrality measure of the hub is given by $x^* = \frac{1+3\alpha}{1-3 \alpha^2}$ and the Katz-Bonacich centrality of a peripheral node by $y^*= \frac{1+\alpha}{1-3 \alpha^2}$ for $\alpha < \frac{1}{\sqrt{3}}$.  Notice that $3y^* > x^*$. It is easy to check that no pair of agents has an incentive to deviate: if the star and the hub exchange their quantities, the difference in utilities is given by

\[ (y^*-x^*) 3 y^* + (x^* - y^*) x^* = (x^* - y^*)(x^* - 3 y^*) < 0, \]

Suppose now that three agents deviate, the hubs of the two stars, agents $1$ and $3$ and a peripheral agent, agent $2$. Furthermore, suppose that agents $2$ and $3$ exchange their quantities. The difference in utilities for the coalition of three agents is given by

\[ 2 (x^* - y^*) x^* - 3 y^*(x^*-y^*) = (x^* - y^*)(2 x^* - 3 y^*). \]

Now, whenever $\alpha > \frac{1}{3}$, this difference in utilities is positive. The loss in utilities of agent $3$ (due to the decrease in its quantity from $x^*$ to $y^*$ is more than compensated by the gain of agents $1$ and $2$, who can choose a transfer to convince agent $3$ to participate in the deviating coalition.
Hence there exist intermediate values of $\alpha$, $\frac{1}{3} < \alpha < \frac{1}{\sqrt{3}}$ for which the coalition of three non-adjacent agents has an incentive to deviate even though no two-player coalition has an incentive to deviate.

\subsubsection{Regular oriented tree}

\begin{example}
\label{exdevor}
Consider the 7 agent network of Figure \ref{figortree}.
\end{example}

 In the optimal contract of the planner

\begin{eqnarray*}
x_1^* & = & a + 3 \alpha x_2^*, \\
x_2^* & = & a + \alpha x_1^* + \alpha x_5^*, \\
x_5^* & = & a + \alpha x_2^*
\end{eqnarray*}

\noindent yielding

\[ x_1^* = \frac{1+2\alpha^2+3\alpha}{1- 4 \alpha^2} > x_2^* = \frac{1}{1-4\alpha^2} > x_3^* = \frac{1-\alpha}{1-4\alpha^2}. \]

Now consider a deviation by the non-adjacent pair $(1,5)$. Then the difference in utilities is given by

\[ (x_1^*-x_5^*) x_2^* > 0, \]
\noindent so that the pair has an incentive to deviate at the first-best contract.

\subsection{Richer information structures and mechanisms} 

\noindent {\em Agents know the identity of their neighbors} 

\bigskip

Suppose that agent $i$ knows the identity of the agents he is influenced by, i.e. agents for which $g_{ij}=1$. Going back to example \ref{ex1}, agent $1$ now observes the identity of his two neighbors and has an information partition cell $E_1(g_1) = \{g_1\}$. All other agents, and the planner, have the same information structure. Under this assumption, the planner can increase the set of possible messages, and ask agents to report an information cell $E$ in a larger set of possible information cells, consistent with the network architecture $\hat{g}$. We show that the planner can exploit the correlation between the announcements of agents to implement the first-best contract, in the spirit of Cr\'{e}mer and McLean (1988)\cite{cremer1988full}

\begin{proposition} \label{propneighbor} Suppose that all agents know the identity of their neighbors. Then there exist a mechanism such that the optimal contract ${\bf x^*} $is  group incentive compatible with transfers.
\end{proposition}

\noindent{\bf Proof:} Consider a contract where the planner assigns $\{x^*_1,...,x^*_n\}$ to agents in location $l=1,..,n$ when announcements are consistent and $\{0,..,0\}$ if announcements are inconsistent. Suppose that there exists a profitable deviation by some coalition $S$ resulting in decisions ${\bf x} \neq {\bf x^*}$. Then clearly announcements must be consistent. By definition, the Because ${\bf x^*}$ maximizes the sum of utilities of the agents, if $\sum_{i \in S} U_i({\bf x}) > \sum_{i \in S} U_i({\bf x^*}$, there must exist an agent $j \notin S$ such that $U_j({\bf x}) < U_j({\bf x^*}$. As announcements are consistent and $j$ truthfully announces his location, he must receive $x^*_j$. Hence if $U_j({\bf x}) < U_j({\bf x^*}$, $\sum_{k} g_{jk} x_k \neq \sum_{k} g_{jk} x_k^*$. This means that the set of agents who declare to be at locations $k$ such that $g_{jk} = 1$ cannot be the true set of neighbors of $j$. But then, there must be an inconsistency in the announcements, contradicting the fact that ${\bf x} \neq \{0,..,0\}$. \hfill $\square$

Proposition \ref{propneighbor} shows that, if agents announce both their location and their list of neighbors, the planner can implement the optimal contract at no cost. This mechanism extracts complex information from the agents, checking the consistency of the list of neighbors and locations announced by each agent. If instead one were to restrict attention to {\em simple} contracts, which are only based on announcements of locations, then the analysis of the benchmark case would prevail. Hence we can reinterpret the benchmark case with finer information structure of the agents, as a restriction on the complexity of the contract. It assumes that the contract is a simple sorting mechanism, where agents only announce locations, and obtain the payoff corresponding to their announced location if a simple consistency check (no two agents announce the same location) is satisfied.

\subsection{Partial information}

\noindent{\em The planner knows the identity of the root} 

\bigskip

Consider a 7 agent network, with a three-tier hierarchy with a single root, $1$, two agents in the second tier $2,3$ and four agents in the third tier, $4,5,6,7$. 

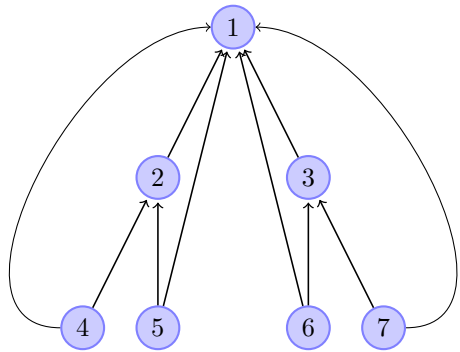
\begin{equation*}
\begin{tikzpicture}[inner sep=1mm]
\tikzstyle{place}=[circle,draw=blue!50,fill=blue!20,thick,minimum size=2mm]
\tikzstyle{post}=[->,shorten >=1pt,semithick]
 \node at (2,2)[place](1){1}; \node at(1,0)[place](2){2} edge[post] node{}(1); \node at (3,0)[place](3){3} edge [post] node{}(1); 
 \node at (0,-2)[place](4){4} edge [post] node{}(2); \node at (1,-2)[place](5){5} edge [post] node{}(2) edge [post] node{}(1) ; 
 \node at (3,-2)[place](6){6} edge [post] node{}(3) edge [post] node{}(1);
 \node at(4,-2)[place](7){7} edge[post] node{}(3); 
 \draw [->] (4) to [out=180,in=180] (1);
 \draw [->] (7) to [out=0,in=0] (1);
  \end{tikzpicture}
\end{equation*}

In the first-best contract, we have

\begin{eqnarray*}
x_1^* & = & a + 2 \alpha x_2^*, \\
x_2^* & = & a + \alpha x_1^* + 2 \alpha x_4^*, \\
x_4^* & = & a + \alpha x_1^* + \alpha x_2^*
\end{eqnarray*}

\noindent resulting in 
\[ x_1^* = \frac{1+2 \alpha + 6 \alpha^2}{1-4 \alpha^2 + 4 \alpha^3}, x_2^* = \frac{1+ 3 \alpha + 2 \alpha^2}{1-4 \alpha^2 + 4 \alpha^3} > x_3^* = \frac{1+2 \alpha + \alpha^2}{1-4 \alpha^2 + 4 \alpha^3}.\]

By Proposition \ref{hiersingle}, if the planner ignores the identity of the root, she must set $x_1 = x_4 = x_5 = x_6$. Now suppose that the planner knows the identity of the root, so that the only possible deviating coalitions are coalitions of agents at tiers $2$ and $3$. The first-best contract assigns $x_4^* < x_2^*$, and the group incentive compatibility constraint writes:

\[ (x_2^*-x_4^*) (x_1^* - x_1^*) \geq 0. \]

\noindent which is always satisfied. Hence the optimal contract satisfies the group incentive compatibility constraint with transfers as soon as the planner knows the identity of the root.

\section{Conclusions}

This paper studies whether a planner who only has information about the network topology can discriminate among agents according to their network position. The planner proposes a simple menu of contracts, one for each location, in order to maximize total welfare, and agents choose among the menu. This mechanism is immune to deviations by single agents, and to deviations by groups of agents of sizes $2$, $3$ and $4$ if side-payments are ruled out. However, if compensations are allowed, groups of agents may have an incentive to jointly deviate from the optimal contract in order to exploit other agents. We identify network topologies for which the optimal contract is group incentive compatible with transfers: undirected networks and regular oriented trees, and network topologies for which the planner must assign uniform quantities: single root and nested neighborhoods directed networks.

The analysis of this paper is a first step in the general study of the possible exploitation of anonymized social data by third parties. However, we realize that the setting we consider is very specific, and further research is needed to understand better how a planner can use partial knowledge of the social network to target agents. In particular, we would like to study how a planner can use simple instruments such as referral fees to extract information about agents' local neighborhoods, without any information about the network topology. 

\newpage

\bibliographystyle{plain}
\bibliography{references}

\appendix 

\section{Proof of proposition \ref{propair} } \label{Appendix1}

Fix a coalition $S$ of adjacent players and consider the permutation $\rho$ mapping $i$ to $\rho(i)$ for any $i \in S$. (We dispense with the index $S$ as the coalition is fixed). We first compute the difference of utility of any player $i$ in $S$ when players in the coalition use the permutation and when they report their true locations for any contract ${\bf g}$, ,

\begin{eqnarray*}
\Delta U_i & = & a(x_{\rho(i)}-x_i) - \frac{(x_i ^2- x_{\rho(i)}^2)}{2}  + \alpha [\sum_{j \in s} g_{ij} (x_{\rho(i)} x_{\rho(j)} - x_i x_j) + \sum_{j \notin s}g_{ij}  (x_{\rho(i)} - x_i) x_j], \\
& = & (x_{\rho(i)}-x_i)[a - \frac{(x_i + x_{\rho(i)})}{2} + \alpha \sum_{j \notin S} g_{ij} x_j] + \alpha \sum_{j \in S} g_{ij} (x_{\rho(i)} x_{\rho(j)} - x_i x_j).
\end{eqnarray*}

At the optimal contract, 

\[ x_i^* = a + \alpha \sum_j [g_{ij}+ g_{ji}] x_j^*.\]

\noindent So

\begin{eqnarray*}
 x_i^* + x_{\rho(i)}^* &= &  2 a + \alpha \sum_j ( [g_{ij}+ g_{ji}] x_j^* + [g_{\rho(i)j} + g_{j\rho(i)}] x_j^*.\\
 & = & 2a + 2 \alpha \sum_j [g_{ij}+ g_{ji}] x_j^* + \alpha ( \sum_j [g_{\rho(i)j} + g_{j\rho(i)}] x_j^* - \sum_j ( [g_{ij}+ g_{ji}] x_j^*))), \\
 & = & 2a + 2 \alpha \sum_j  [g_{ij}+ g_{ji}] x_j^* + (x_{\rho(i)}^* - x_i)
 \end{eqnarray*}
 
 \noindent Replacing we obtain

\begin{eqnarray*}
\Delta U_i & = &  \alpha[ (x^*_{\rho(i)}-x^*_i) ( - \sum_{j \notin S} g_{ji} x_j^* + \frac{(x_{\rho(i)}^* - x^*_i)}{2} ) \\
& +  & \sum_{j \in S} (g_{ij} (x^*_{\rho(i)} x^*_{\rho(j)} - x^*_i x^*_j -  x^*_j (x^*_{\rho(i)}-x_i^*))) - g_{ji} x^*_j (x^*_{\rho(i)}-x_i^*) )] \\
 & = &  \alpha [- \sum_{j \notin S} (x^*_{\rho(i)}-x^*_i) g_{ji} x_j^* - \frac{(x^*_{\rho(i)}-x^*_i)^2}{2} + \sum_{j \in S} g_{ij} x^*_{\rho(i)}(x^*_{\rho(j)}-x^*_j) - \sum_{j \in S} g_{ji} x^*_j (x^*_{\rho(i)}-x_i^*)].
\end{eqnarray*}

\noindent Hence $\Delta U_i >0$ if and only if

\begin{equation}
\label{ineqpair}
\sum_{j \in S} g_{ij} x^*_{\rho(i)}(x^*_{\rho(j)}-x_j) - \sum_{j \in S} g_{ji} x^*_j (x^*_{\rho(i)}-x_i^*) >   \sum_{j \notin S} (x^*_{\rho(i)}-x^*_i) g_{ji} x_j^* + \frac{(x^*_{\rho(i)}-x^*_i)^2}{2}.
\end{equation}

\noindent We now show that when $|S| \leq 4$, there must be some agent for whom inequality \eqref{ineqpair} fails. 

Suppose first that $|S|=2$. Suppose without loss of generality that $x^*_2 > x^*_1$ and consider $\Delta U_1$. Inequality \eqref{ineqpair} implies that

\[ \Delta U_1 =  g_{12}+ g_{21} x^*_2 (x^*_2 - x^*_1) >0, \]

\noindent a contradiction as $x^*_2 > x^*_1$.

Suppose next that $|S|=3$ and suppose without loss of generality that $x^*_1 < x^*_2 < x^*_3$. Agent $1$ must trade his quantity $x^*_1$ for a higher quantity. Suppose first that $\rho(1)=3, \rho(3)=2, \rho(2)=1$ so that agent $1$ is the only agent who trades his quantity $x_1^*$ for a higher quantity. Then inequality (\ref{ineqpair}) yields

\[ \Delta U_1 = g_{12} x^*_3 (x^*_1-x^*_2) + g_{13}x^*_3(x^*_2 - x^*_3) - x^*_2(x^*_2-x^*_1) - x^*_3(x^*_2-x^*_1) >0, \]

\noindent a contradiction as $x^*_3 > x^*_2 > x^*_1$.

Suppose now that $\rho(1)=2, \rho(2)=3$ and $\rho(3)=1$. (This is the only other permutation to consider, as any other permutation on $S$ will leave one of the players at the same position.) Then inequalities (\ref{ineqpair}) for agents $1$ and $2$ result in

\begin{eqnarray*}
\Delta U_1 = g_{12} x^*_2(x^*_3-x^*_2) + g_{13} x^*_2(x^*_1-x^*_3) - g_{21} x^*_2(x^*_2-x^*_1) - g_{31} x^*_3(x^*_2-x^*_1) & > & 0, \\
\Delta U_2 = g_{21} x^*_3 (x^*_2-x^*_1) + g_{23} x^*_3(x^*_1-x^*_3) - g_{12} x^*_1(x^*_3-x^*_2) - g_{32} x^*_3(x^*_3-x^*_2) & > & 0.
\end{eqnarray*}

First notice that in both inequalities, the only positive term is the first term, which implies that both $g_{12}$ and $g_{21}$ must be equal to $1$ for the two inequalities to hold. Second, recall, that because the coalition is connected $g_{13}+g_{31} \geq 1$ and $g_{23} + g_{32} \geq 1$. Now, as $x^*_2(x^*_3-x^*_1) > x^*_2(x^*_2 - x^*_1)$, the first inequality is more likely to be satisfied when $g_{13}=0$ and $g_{31}=1$. Similarly, because $x^*_3(x^*_3-x^*_1) > x^*_3(x^*_3-x^*_2)$, the second inequality is more likely to be satisfied when $g_{23}=0$ and $g_{32=1}$. Hence if both inequalities are satisfied, we must have

\begin{eqnarray*}
x^*_2(x^*_3-x^*_2)  & > & x^*_2(x^*_2-x^*_1) + x^*_3(x^*_2-x^*_1), \\
x^*_3 (x^*_2-x^*_1) & > & x^*_1(x^*_3-x^*_2) +  x^*_3(x^*_3-x^*_2)
\end{eqnarray*}

But this implies

\[ x^*_2(x^*_3-x^*_2) >  x^*_2(x^*_2-x^*_1) + x^*_3(x^*_2-x^*_1) > x^*_2(x^*_2-x^*_1) + x^*_1(x^*_3-x^*_2) +  x^*_3(x^*_3-x^*_2) > x^*_3(x^*_3-x^*_2), \]

\noindent a contradiction that completes the proof for $|S|=3$ .

Now suppose that the size of the group is $|S|=4$ and suppose without loss of generality that $x_4^{*} > x_3^{*} > x_2^{*} > x_1^{*}$. There are six possible permutations where each member of the group gets a different contract from the one intended for their location and no pair exchanges pairwise their contracts. There are $3^6$ possible $ways$ in which the members of the group can be linked because each link $(g_{ij}, g_{ji}) \in   \{ (0,1) , (1,0) , (1,1)\}$. \\
\\ Recall that : 
\begin{equation*}
\begin{split}
\Delta U_i =  - \frac{\alpha}{2} (x_{\rho(i)} - x_i)^2 & - \alpha (x_{\rho(i)} - x_i) \sum_{j \in {S}} g_{ji} x_ j  \\
& - \alpha (x_{\rho(i)} - x_i) \sum_{j \not \in {S}} g_{ji} x_ j \\
& + \alpha x_{\rho(i)} \sum_{j \in {S}} g_{ij} (x_{\rho(j)} -x_j) 
\end{split}
\end{equation*}
\begin{enumerate}
\item Consider: $\rho(1)=4$, $\rho(2)=1$, $\rho(3)=2$ and $\rho(4)=3$  \\ \\ Note that all agents get a lower quantity except agent $1$. The difference in utility of agent $1$ writes: 
\begin{equation*}
	\begin{split}
		\Delta U_1 =  - \frac{\alpha}{2} (x_4^{*} - x_1^{*})^2 & - \alpha {(x_4^{*} - x_1^{*})} [g_{21} x_ 2  + g_{31} x_ 3 + g_{41} x_ 4] - \alpha {(x_4^{*} - x_1^{*})} \sum_{j \not \in {S}} g_{j1} x_ j \\
& + \alpha x_{\rho(1)}[ g_{12} {(x_1^{*} -x_2^{*})} + g_{13} {(x_2^{*} -x_3^{*})} + g_{14} {(x_3^{*} -x_4^{*})}] < 0
	\end{split}
\end{equation*}

\noindent  Recall that $x_1^{*} < x_2^{*} < x_3^{*} < x_4^{*}$, hence there exists at least one agent that is strictly worse off with $\rho$ for all network structures i.e. for all $i \neq j \in \{ 1,2,3,4\}$ and $ g_{ij} \in \{ 0,1\}$ such that $g_{ij}+ g_{ji} \geq 1$. \\ \\ 
 \item Consider: $\rho(1)=4$, $\rho(2)=3$, $\rho(3)=1$ and $\rho(4)=2$. 
\\ \\ For all the network structures such that $g_{12}=0$ and the remaining links take the value $0$ or $1$ such that for all $i , j \in S$ $g_{ij}+g_{ji} \geq 1$,  $\Delta U_1<0$. Similarly, for $g_{21}=0$,  $\Delta U_2<0$. 
\\ \\  Moreover, when $g_{12}=g_{21}=1$ and $g_{13}=1$ and $g_{31} \in \{ 0,1\}$: $\Delta U_1<0$. For $g_{12}=g_{21}=1$ and $g_{14}=1$ and $g_{41} \in \{ 0,1\}$: $\Delta U_1<0$.  The remaining networks that we need inspect are networks such that $g_{12}=g_{21}=1$ and $g_{13}=0$, $g_{31}=1$, $g_{14}=0$, $g_{41}=1$. But again, when $g_{31}=1$ and the remaining links take values $0$ or $1$ such that for $i, j \in S$, $g_{ij} + g_{ji} \geq 1$, we get $\Delta U_1 < 0$.

   \item Consider: $\mathbf{\rho(1)=3}$, $\mathbf{\rho(2)=4}$, $\rho(3)=2$ and $\rho(4)=1$. \\ \\ For all the network structures such that $g_{12}=0$ and the remaining links take the value $0$ or $1$ such that for all $i , j \in S$ $g_{ij}+g_{ji} \geq 1$,  $\Delta U_1<0$. Similarly, when $g_{21}=0$, $\Delta U_2<0$. Hence we check structures where $g_{12}=g_{21}=1$.  \\ \\
For $g_{12}=g_{21}=1$ and $g_{24}=1$ and $g_{42} \in \{ 0,1\}$: $\Delta U_2<0$. Similarly, when $g_{12}=g_{21}=1$ and $g_{14}=1$ and $g_{41} \in \{ 0,1\}$: $\Delta U_1<0$. 		
\\ \\ The remaining networks that we need to inspect are such that {$g_{12}=g_{21}=1$ and $g_{14}=0$, $g_{41}=1$, $g_{24}=0$, $g_{42}=1$} and the remaining links take the value $0$ or $1$, and for all $i , j \in S$ $g_{ij}+g_{ji} \geq 1$. Consider $\Delta U_2$ : 
\begin{equation*} 
	\begin{split}
		\Delta U_2 & = - \frac{\alpha}{2} (x_4^{*} - x_2^{*})^2 - \alpha (x_4^{*} - x_2^{*}) \sum_{j \not \in S } g_{j2} x_j 
		 - \alpha (x_4^{*} - x_2^{*}) ( x_1^{*} + g_{32} x_3^{*} + x_4^{*}) \\
		& + \alpha x_4^{*} ( (x_3^{*} - x_1^{*}) + g_{23} (x_2^{*} - x_3^{*}) )) 
	\end{split}
\end{equation*}

If $(x_3^{*} - x_1^{*}) < ( x_4^{*} - x_2^{*})$ then $\Delta U_2 < 0$. If not, consider $\Delta U_1$: 
\begin{equation*} 
	\begin{split}
		\Delta U_1 & = - \frac{\alpha}{2} (x_3^{*} - x_1^{*})^2 - \alpha (x_3^{*} - x_1^{*}) \sum_{j \not \in S } g_{j1} x_j 
		 - \alpha (x_3^{*} - x_1^{*}) (  x_2^{*} + g_{31} x_3^{*} +  x_4^{*}) \\
		& + \alpha x_3^{*} (  (x_4^{*} - x_2^{*}) + g_{13} (x_2^{*} - x_3^{*})  )) 
	\end{split}
\end{equation*}
If $(x_3^{*} - x_1^{*}) \geq ( x_4^{*} - x_2^{*})$ then $\Delta U_1 <0$. A contradiction which completes the proof for this permutation $\rho$.

  \item Consider: $\mathbf{\rho(1)=3}$, $\rho(2)=1$, $\mathbf{\rho(3)=4}$ and $\rho(4)=2$. \\ \\
 For all the network structures such that $g_{31}=0$ and the remaining links take the value $0$ or $1$ such that for all $i , j \in S$ $g_{ij}+g_{ji} \geq 1$,  $\Delta U_3<0$. Similarly, when $g_{13}=0$, $\Delta U_1<0$. Moreover for $g_{13}=g_{31}=1$ and $g_{14}=1$ and $g_{41} \in \{ 0,1\}$ , it follows that $\Delta U_1<0$. \\ \\
The remaining networks to inspect are ones where $g_{13}=g_{31}=1$ and  $g_{14}=0$ and $g_{41}=1$. Let  $ d_1 = x_3^{*} - x_1^{*} $ and $ d_2 = x_4^{*} - x_3^{*} $. The variation in payoffs of agents $1$ and $3$ writes: 
\begin{equation*}
\begin{split}
\Delta U_1 = & - {\alpha} d_1 [\frac{d_1}{2} + (g_{21} x_2^{*} + x_3^{*} + x_4^{*} ) + \sum_{j \not \in S} g_{j1} x_j ] - \alpha x_3^{*} g_{12} (x_2^{*} - x_1^{*}) + \alpha d_2 x_3^{*}  \\
\Delta U_3 = & - {\alpha} d_2 [\frac{d_2}{2} + (g_{23} x_2^{*} + x_1^{*} + g_{43} x_4^{*} ) + \sum_{j \not \in S} g_{j3} x_j ] - \alpha x_4^{*} [g_{32} (x_2^{*} - x_1^{*}) + g_{34} (x_4^{*} - x_2^{*})] + \alpha d_1 x_4^{*} 
\end{split}
\end{equation*}
If $d_1 \geq d_2$ then $\Delta U_1 <0$ because the positive term $ \alpha d_2 x_3^{*}$ is strictly smaller than $ - \alpha d_1 x_4^{*}$. We study below the remaining case: $d_2 > d_1$. Suppose by contradiction that $\Delta U_3 >0$ and recall that $g_{34} + g_{43} \geq 1$ because players $3$ and $4$ are part of the same group $S$. For $g_{43}=1$ and $g_{34} \in \{ 0,1\}$:

\begin{equation*}
\begin{split}
\Delta U_3 > 0  &  \Leftrightarrow \alpha d_1 x_4^{*} > {\alpha} d_2 [\frac{d_2}{2} + (g_{23} x_2^{*} + x_1^{*} +  x_4^{*} ) + \sum_{j \not \in S} g_{j3} x_j ] + \alpha x_4^{*} [g_{32} (x_2^{*} - x_1^{*}) + g_{34} (x_4^{*} - x_2^{*})] {( > 0 )}  \\
 & \Leftrightarrow \alpha x_4^{*} (d_1 - d_2 )  > {\alpha} d_2 [\frac{d_2}{2} + (g_{23} x_2^{*} + x_1^{*}  ) + \sum_{j \not \in S} g_{j3} x_j ] + \alpha x_4^{*} [g_{32} (x_2^{*} - x_1^{*}) + g_{34} (x_4^{*} - x_2^{*})] {(> 0 )}
\end{split}
\end{equation*}
A contradiction because $\alpha x_4^{*} (d_1 - d_2 ) < 0$. Hence $\Delta U_3 <0$.  Recall that $\alpha d_1 x_4^{*}$ is upper bounded by $\alpha d_2 x_4^{*}$ because we are considering the case $d_2 > d_1$. For $g_{34}=1$ and $g_{43} \in \{ 0,1\}$:
\begin{equation*}
\begin{split}
 & \alpha d_2 x_4^{*} > ( \alpha d_1 x_4^{*} > ) {\alpha} d_2 [\frac{d_2}{2} + (g_{23} x_2^{*} + x_1^{*} + { g_{43} x_4^{*}} ) + \sum_{j \not \in S} g_{j3} x_j ] + \alpha x_4^{*} [g_{32} (x_2^{*} - x_1^{*}) + { (x_4^{*} - x_2^{*})}]  {(> 0 )} \\
\Leftrightarrow & \alpha d_2 x_4^{*} - \alpha x_4^{*} (x_4^{*} - x_2^{*})  > {\alpha} d_2 [\frac{d_2}{2} + (g_{23} x_2^{*} + x_1^{*}  ) + \sum_{j \not \in S} g_{j3} x_j ] + \alpha x_4^{*} g_{32} (x_2^{*} - x_1^{*}) {( > 0 )} \\
\Leftrightarrow & \alpha x_4^{*} ( x_4^{*} - x_3^{*}- x_4^{*} + x_2^{*})  > {\alpha} d_2 [\frac{d_2}{2} + (g_{23} x_2^{*} + x_1^{*}  ) + \sum_{j \not \in S} g_{j3} x_j ] + \alpha x_4^{*} g_{32} (x_2^{*} - x_1^{*})  { ( > 0 )} \\ 
\end{split}
\end{equation*}
A contradiction because $x_1^{*} < x_2^{*} < x_3^{*} < x_4^{*}$ implies $\alpha x_4^{*} ( x_4^{*} - x_3^{*}- x_4^{*} + x_2^{*}) < 0$ and the $RHS$ is strictly positive. Hence $\Delta U_3 <0$, which completes the proof for this permutation.

    \item Consider: $\rho(1)=2$, $\mathbf{\rho(2)=4}$, $\rho(3)=1$ and $\mathbf{\rho(4)=3}$. \\ \\
 For all the network structures such that $g_{21}=0$ (link with the other agent who gets a higher quantity) and the remaining links take the value $0$ or $1$ such that for all $i , j \in S$ $g_{ij}+g_{ji} \geq 1$,  $\Delta U_2<0$.  Similarly, for $g_{12}=0$, $\Delta U_1<0$. Moreover, for $g_{12}=g_{21}=1$ and $g_{23}=1$ and $g_{32} \in \{ 0,1\}$: $\Delta U_2<0$.
We inspect the remaining group structures for which we did not show a contradiction: $g_{12}=g_{21}=1$ and  $g_{23}=0$ and $g_{32}=1$. Following the same reasoning in the previous case, let $d_3=x_2^{*} - x_1^{*}$ and $d_4 = x_4^{*} - x_2^{*}$. If $d_3 \geq d_4=$ then $\Delta U_1 <0$. If $d_4 > d_3$ then $\Delta U_3 <0$ (consider the links $g_{42}+g_{24} \geq 1$). 
		
		 \item Consider: $\rho(1)=2$, $\rho(2)=3$, $\rho(3)=4$ and $\rho(4)=1$  \\ \\ \noindent Let $d_1=x_2^{*} - x_1^{*}$, $d_2=x_3^{*} - x_2^{*}$, $d_3 = x_4^{*} - x_3^{*}$ and $d_4=x_4^{*} - x_1^{*}$ (clearly $d_4 = d_1 + d_2 + d_3$). The variation in payoffs of agents $1$, $2$, $3$ write: 
		\begin{equation*}
			\begin{split}
				\Delta U_1 & = - \alpha d_1 [\frac{d_1}{2} + g_{21} x_2^{*} + g_{31} x_3^{*} + g_{41} x_4^{*} + \sum_{j \not \in S} g_{j1} x_j ] 
						  + \alpha x_2^{*} [ g_{12} d_2 + g_{13} d_3 - g_{14} d_4]
			\end{split}
		\end{equation*}
		If $g_{14} =1$, $g_{12} \in \{ 0,1\}$ and $g_{13} \in \{ 0,1\}$ then $\Delta U_1 <0$ because $d_4 > d_2 + d_3$. If $g_{12} = g_{13} =0$ and $g_{14} \in \{ 0,1\}$ then $\Delta U_1 <0$. Hence we consider networks where $g_{14}=0$ and $g_{12} + g_{13} \geq 1$. 
			\begin{equation*}
			\begin{split}
				\Delta U_2 & = - \alpha d_2 [\frac{d_2}{2} + g_{12} x_1^{*} + g_{32} x_3^{*} + g_{42} x_4^{*} + \sum_{j \not \in S} g_{j2} x_j ] 
						  + \alpha x_3^{*} [ g_{21} d_1 + g_{23} d_3 - g_{24} d_4]
			\end{split}
		\end{equation*}
		If $g_{24} =1$, $g_{21} \in \{ 0,1\}$ and $g_{23} \in \{ 0,1\}$ then $\Delta U_2 <0$. If $g_{21} = g_{23} =0$ and $g_{24} \in \{ 0,1\}$ then $\Delta U_2 <0$. Hence we consider networks where $g_{24}=0$ and $g_{21} + g_{23} \geq 1$.  
				\begin{equation*}
			\begin{split}
				\Delta U_3 & = - \alpha d_3 [\frac{d_3}{2} + g_{13} x_1^{*} + g_{23} x_2^{*} + g_{43} x_4^{*} + \sum_{j \not \in S} g_{j3} x_j ] 
						  + \alpha x_4^{*} [ g_{31} d_1 + g_{32} d_2 - g_{34} d_4]
			\end{split}
		\end{equation*}
		 If $g_{34} =1$, $g_{31} \in \{ 0,1\}$ and $g_{32} \in \{ 0,1\}$ then $\Delta U_3 <0$. If $g_{31} = g_{32} =0$ and $g_{34} \in \{ 0,1\}$ then $\Delta U_3 <0$. Hence we consider networks where $g_{34}=0$ and $g_{31} + g_{32} \geq 1$. \\ \\
		 Now we consider the remaining group structures such that: $g_{14}=g_{24}=g_{34}=0$ (and $g_{41}=g_{42}=g_{43}=1$), $g_{12} + g_{13} \geq 1$, $g_{21} + g_{23} \geq 1$ and $g_{31} + g_{32} \geq 1$. In addition to the three previous inequalities, $g_{ij} + g_{ji} \geq 1$ for $i \neq j \in S$. This leaves us with exactly $10$ group structures to inspect. The variation in payoffs rewrite: 		 		\begin{equation*}
			\begin{split}
				\Delta U_1 & = - \alpha d_1 [\frac{d_1}{2} + g_{21} x_2^{*} + g_{31} x_3^{*}   + \sum_{j \not \in S} g_{j1} x_j ] - \alpha d_1 x_4^{*}
						  + \alpha x_2^{*} [ g_{12} d_2 + g_{13} d_3 ] \\
				\Delta U_2 & = - \alpha d_2 [\frac{d_2}{2} + g_{12} x_1^{*} + g_{32} x_3^{*}  + \sum_{j \not \in S} g_{j2} x_j ] - \alpha d_2 x_4^{*}
						  + \alpha x_3^{*} [ g_{21} d_1 + g_{23} d_3 ] \\
				\Delta U_3 & = - \alpha d_3 [\frac{d_3}{2} + g_{13} x_1^{*} + g_{23} x_2^{*}  + \sum_{j \not \in S} g_{j3} x_j ]  - \alpha d_3 x_4^{*}
						  + \alpha x_4^{*} [ g_{31} d_1 + g_{32} d_2]
			\end{split}
		\end{equation*}

		\begin{enumerate} 
		\item $g_{12} =g_{31} =g_{23}=1 $ and $g_{13} =g_{32} =g_{21}=0 $: if $d_3 > d_1$ then $\Delta U_3 <0$. If $d_3 < d_1$ and $d_2 > d_3$ then $\Delta U_2 <0$. If $d_3 < d_1$ and $d_3 > d_2$ then it implies $d_1 > d_2$ and $\Delta U_1 <0$.  \\
		\item  $g_{12} =g_{31} =g_{23}=0 $ and $g_{13} =g_{32} =g_{21}=1 $: if $d_3 > d_2$ then $\Delta U_3 <0$. If $d_2 < d_3$ and $d_2 > d_1$, $\Delta U_2 <0$. If $d_2 > d_3$ and $d_1 > d_2$ then it implies $d_1 > d_3$, $\Delta U_1 <0$. \\
		\item  $g_{12} =g_{31} =g_{32}= g_{23} =g_{21} =1$ and $g_{13}=0 $: if $d_1 > d_2$, $\Delta U_1 <0$. If $d_2 > d_1$, $\Delta U_2 <0$.  \\
		\item   $g_{13} =g_{31} =g_{23}= g_{21} =1$ and $g_{12}= g_{32} = 0 $: if $d_1 > d_3$, $\Delta U_1 <0$. If $d_3 > d_1$, $\Delta U_3 <0$.  \\
		\item $g_{12} =g_{13} =g_{32}= g_{23} =1$ and $g_{31}= g_{21} = 0 $: if $d_2 > d_3$, $\Delta U_2 <0$. If $d_3 > d_2$, $\Delta U_3 <0$.  \\
		\item  $g_{12} =g_{13} =g_{32}= g_{23} = g_{21} = 1$ and $g_{31} = 0$: if $d_1 > d_2 $ and $d_1 > d_3$ then $\Delta U_1 <0$. If $d_1 > d_2 $ and $d_3 > d_1$ then it implies $d_3 > d_2$ and $\Delta U_3 <0$. If $d_2 > d_1 $ and $d_1 > d_3$ then it implies $d_2 > d_3$ and $\Delta U_2 <0$. If $d_2 > d_1 $ and $d_3 > d_1$ then $\Delta U_3 <0$.  \\
		\item $g_{12} =g_{13} =g_{31}= g_{32} = g_{21} = 1$ and $g_{23} = 0$. If $d_2 > d_1$ then $\Delta U_2 < 0$. If $d_1 > d_2$ and $d_1 > d_3$ then $\Delta U_3 < 0$. If $d_1 > d_2$ and $d_3 > d_1$, it implies $d_3 > d_1 > d_2$. If $d_3 > d_1 > d_2$ and $d_3 > d_1 + d_2 $ then $\Delta U_3 < 0$. Last case: $d_3  > d_1 > d_2 $ and $ d_3 < d_1 + d_2$.  We rewrite $\Delta U_1$ and find a (negative) upper bound using $d_3 < d_1 + d_2 \Leftrightarrow d_3 - d_1 < d_2$: 
		\begin{equation*}
		\begin{split}
			& \Delta U_1 = - \alpha d_1 (\frac{d_1}{2} + \sum_{j \not \in S} g_{j1} x_j) - \alpha d_1 x_3^{*} + \alpha d_2 x_2^{*}  - \alpha d_1 x_4^{*} + \alpha x_2^{*} (d_3 - d_1) \\
			& <  - \alpha d_1 (\frac{d_1}{2} + \sum_{j \not \in S} g_{j1} x_j) - \alpha d_1 x_3^{*} + \alpha d_2 x_2^{*}  - \alpha d_1 x_4^{*} + \alpha x_2^{*} d_2 \\
			& < 0 \text{ because } d_1 > d_2 
		\end{split}
		\end{equation*}
		\item $g_{12} =g_{13} =g_{23}= g_{32} = g_{21} = 1$ and $g_{31} = 0$. If $d_1 > d_2$ and $d_1 > d_3$ then $\Delta U_1<0$. If $d_1 > d_2$ and $d_3 > d_1$ then $d_3 > d_2$ and $\Delta U_3<0$. If $d_2 > d_1$ and $d_1 > d_3$ then $d_2 > d_3$ and $\Delta U_2<0$. {Last two cases}: if $d_2 > d_1$ and $d_3 > d_1$ and $d_2 >d_3$. We rewrite $\Delta U_2$: 
		\begin{equation*}
		\begin{split}
			\Delta U_2 = - \alpha d_2 (\frac{d_2}{2} + \sum_{j \not \in S} g_{j2} x_j) - d_2 x_1^{*} - x_3^{*} (d_2 - d_3 ) - d_2 x_4^{*} +d_1 x_3^{*} < 0 
		\end{split}
		\end{equation*}
		If $d_2 > d_1$ and $d_3 > d_1$ and $d_3 >d_2$, then $\Delta U_3 < 0$ (because here $g_{32}=0$).\\
		
		\item $g_{31}  = g_{12} = g_{21}=g_{23} = 1$ and $g_{32} = g_{13}=0$. If $d_3 > d_1$ then $\Delta U_3 < 0$. If $d_1 > d_3$ and $d_1 > d_2$ then $\Delta U_1 < 0$. {Last two cases}: if $d_2 > d_1 > d_3$ and $d_2 > d_1 + d_3$ then  $\Delta U_2 < 0$. If $d_2 > d_1 > d_3$ and $d_2 < d_1 + d_3$, we rewrite $\Delta U_1$: 
	\begin{equation*}
			\begin{split}
				\Delta U_1 & = - \alpha d_1 [\frac{d_1}{2} + \sum_{j \not \in S} g_{j1} x_j ] - \alpha d_1 (x_2^{*} + x_3^{*} + x_4^{*} ) + \alpha d_2 x_2^{*}    \\
				& < - \alpha d_1 [\frac{d_1}{2} + \sum_{j \not \in S} g_{j1} x_j ] - \alpha d_1 (x_2^{*} + x_3^{*} + x_2^{*} ) + \alpha d_2 x_2^{*}   (\mbox{ because } - x_4^{*} < - x_2^{*}) \\
				& = - \alpha d_1 [\frac{d_1}{2} + \sum_{j \not \in S} g_{j1} x_j ] - \alpha d_1x_3^{*}  - 2 \alpha d_1   x_2^{*}  + \alpha x_2^{*}  d_2  \\
				& < 0 \mbox{ because } (d_2 < d_1 + d_3 < 2 d_1)  
		\end{split}
		\end{equation*}	
		
	\item $g_{31} =g_{13} =g_{32}= g_{23} = g_{21} = 1$ and $g_{12} = 0$. If $d_1 > d_3$ then $\Delta U_1 < 0$. If $d_3 > d_1$ and $d_2 > d_3$ then $\Delta U_2 < 0$. {Last two cases}: if $d_3 > d_2 > d_1$ and $d_3 > d_1 + d_2$ then $\Delta U_3 < 0$. If $d_3 > d_1 > d_2$  and $d_3 < d_1 + d_2$, we rewrite $\Delta U_1$: 
	\begin{equation*}
		\begin{split}
			\Delta U_1 & = - \alpha d_1 (\frac{d_1}{2} + \sum_{j \not \in S} g_{j1} x_j) - \alpha d_1 ( x_2^{*} + x_3^{*} + x_4^{*} ) + \alpha d_3 x_2^{*} \\
			& < - \alpha d_1 (\frac{d_1}{2} + \sum_{j \not \in S} g_{j1} x_j) - \alpha d_1 ( x_2^{*} + x_3^{*} + x_2^{*} ) + \alpha d_3 x_2^{*} \mbox{ because } - x_4^{*} < - x_2^{*}  \\
			& = - \alpha d_1 (\frac{d_1}{2} + \sum_{j \not \in S} g_{j1} x_j) - \alpha d_1 x_3^{*} - 2 \alpha d_1 x_2^{*} + \alpha d_3 x_2^{*} \\
			& < 0 \mbox{ because } d_3 < d_1 + d_2 < 2 d_1 
		\end{split}
	\end{equation*}
		
		\end{enumerate}
\end{enumerate}

\noindent a contradiction that completes the proof of the Proposition. \hfill $\square$.

\end{document}